\documentclass[pdflatex,sn-mathphys-num]{sn-jnl}

\usepackage{array}
\usepackage{booktabs}
\usepackage{lineno,hyperref}
\usepackage{subfigure}
\usepackage{color}
\usepackage{longtable}
\usepackage{caption}
\usepackage{multirow}
\usepackage{hyperref}
\usepackage{enumitem}
\usepackage{svg}
\usepackage{colortbl}
\usepackage{graphicx}%
\usepackage{multirow}%
\usepackage{amsmath,amssymb,amsfonts}%
\usepackage{amsthm}%
\usepackage{mathrsfs}%
\usepackage[title]{appendix}%
\usepackage{xcolor}%
\usepackage{textcomp}%
\usepackage{manyfoot}%
\usepackage{booktabs}%
\usepackage{algpseudocode}%
\usepackage{listings}%
\usepackage{float}

\theoremstyle{thmstyleone}%
\theoremstyle{thmstyletwo}%

\theoremstyle{thmstylethree}%

\begin{document}

\title[Article Title]{A Bioinformatic Approach Validated Utilizing Machine Learning Algorithms to Identify Relevant Biomarkers and Crucial Pathways in Gallbladder 
Cancer}


\author[1,2]{\fnm{Rabea} \sur{Khatun}}\email{rabeakhatun650@gmail.com}

\author[2]{\fnm{Wahia} \sur{Tasnim}}\email{wahiatasnim@gmail.com}

\author[1]{\fnm{Maksuda} \sur{Akter}}\email{maksudaoni6@gmail.com}

\author*[1]{\fnm{Md Manowarul} \sur{ Islam}}\email{manowar@cse.jnu.ac.bd}

\author[1]{\fnm{Dr. Md. Ashraf } \sur{ Uddin}}\email{ashraf@cse.jnu.ac.bd}

\author[1]{\fnm{Dr. Md. Zulfiker} \sur{  Mahmud}}\email{zulfiker@cse.jnu.ac.bd}

\author[1,2]{\fnm{Saurav} \sur{ Chandra Das}}\email{sauravchandradas@gmail.com}

\affil*[1]{\orgdiv{Department of Computer Science and Engineering}, \orgname{Jagannath University}, \orgaddress{\street{9-10 Chittaranjan Ave}, \city{Dhaka}, \postcode{1100},\state{} \country{Bangladesh}}}

\affil[2]{\orgdiv{Department of Computer Science and Engineering}, \orgname{Green University of Bangladesh}, \orgaddress{\street{Purbachal American City}, \city{Kanchan, Rupganj, Narayanganj}, \postcode{1461}, \state{Dhaka}, \country{Bangladesh}}}


\abstract{
Gallbladder cancer (GBC) is the most frequent cause of disease among biliary tract neoplasms. Identifying the molecular mechanisms and biomarkers linked to GBC progression has been a significant challenge in scientific research. Few recent studies have explored the roles of biomarkers in GBC. Our study aimed to identify biomarkers in GBC using machine learning (ML) and bioinformatics techniques.
We compared GBC tumor samples with normal samples to identify differentially expressed genes (DEGs) from two microarray datasets (GSE100363, GSE139682) obtained from the NCBI GEO database. A total of 146 DEGs were found, with 39 up-regulated and 107 down-regulated genes. Functional enrichment analysis of these DEGs was performed using Gene Ontology (GO) terms and REACTOME pathways through DAVID. The protein-protein interaction network was constructed using the STRING database.
To identify hub genes, we applied three ranking algorithms: Degree, MNC, and Closeness Centrality. The intersection of hub genes from these algorithms yielded 11 hub genes. Simultaneously, two feature selection methods (Pearson correlation and recursive feature elimination) were used to identify significant gene subsets. We then developed ML models using SVM and RF on the GSE100363 dataset, with validation on GSE139682, to determine the gene subset that best distinguishes GBC samples. The hub genes outperformed the other gene subsets. Finally, NTRK2, COL14A1, SCN4B, ATP1A2, SLC17A7, SLIT3, COL7A1, CLDN4, CLEC3B, ADCYAP1R1, and MFAP4 were identified as crucial genes, with SLIT3, COL7A1, and CLDN4 being strongly linked to GBC development and prediction.
}




\maketitle
\section{Introduction}\label{sec1}

The most prevalent form of Biliary Tract Cancer (BTC), Gallbladder Cancer (GBC) has an unfavourable outcome and a high death rate\citep{hundal2014gallbladder}\citep{rakic2014gallbladder}\citep{siegel2021cancer}\citep{valle2021biliary}. With an average survival of less than six months as well as a total five-year survival rate of under five percent, this malignancy is an extremely deadly illness. Since this cancer spreads quietly when a late diagnosis is made, early detection is crucial. With an ordinary cholecystectomy in suspicious gallbladder stone illness, 0.5–1.5\% of patients were found with gallbladder cancer \citep{gourgiotis2008gallbladder}.
The eighth American Joint Committee on Cancer (AJCC) guideline \citep{chun2018ajcc} states that the most effective possible treatment for GBC at its infancy is surgical resection; for GBC at a later stage, chemotherapy, radiation therapy, immunotherapy, and targeted therapy are advised. The extremely aggressive and metastatic features of advanced GBC, such as local development of tumours, hepatic invasion, and lymph node metastases, result in minimal reaction to treatment and an unfavourable outlook for victims \citep{mantripragada2017adjuvant} \citep{chen2019development}. While the complicated procedure and the molecular mechanism of GBC are ambiguous, numerous investigations have stated the essential function of numerous biological processes in cancer spread and invasion, including immune evasion \citep{tauriello2018tgfbeta}, epithelial-mesenchymal transition (EMT) \citep{wu2019epithelial} \citep{zheng2018elf3}, and cancer stem cells \citep{civenni2019epigenetic}. In order to enhance the prognosis for GBC patients, it is imperative to investigate the novel biomarkers linked to invasion and spreading.


An emerging method in studies on cancer for discovering pathways and genes as potential prognostic and diagnostic biomarkers is transcriptome analysis of high-throughput sequencing, such as microarrays and RNA sequencing \citep{raphael2017integrated} \citep{kinde2013evaluation}. Additionally, bioinformatics are currently employed to uncover biomarkers linked to certain diseases. An advancement in the better prevention and treatment of GBC could be achieved by these biomarkers \citep{kulasingam2008strategies}.  Relevant gene biomarkers for GBC are currently being identified by bioinformatic analysis of gene expression data; yet, the behaviour of the findings of bioinformatics is inconsistent. 
The predictive importance of several DEGs in GBC has been shown by current research. But the outcomes of such research have been inconsistent, maybe because various statistical techniques were employed. Furthermore, there is still a shortage of examination of the predictive significance of the DEGs utilising machine learning technologies in GBC. Moreover, the enrichment pathways, Gene Ontology (GO) functions, and interaction network of DEGs are still unclear. A combination of bioinformatic methodologies and machine learning techniques can yield trustworthy outcomes and improve the GBC biomarker training and verification process\citep{auslander2021incorporating}.

\begin{figure}[]
    \centering
    \includegraphics[scale=.7
    ]{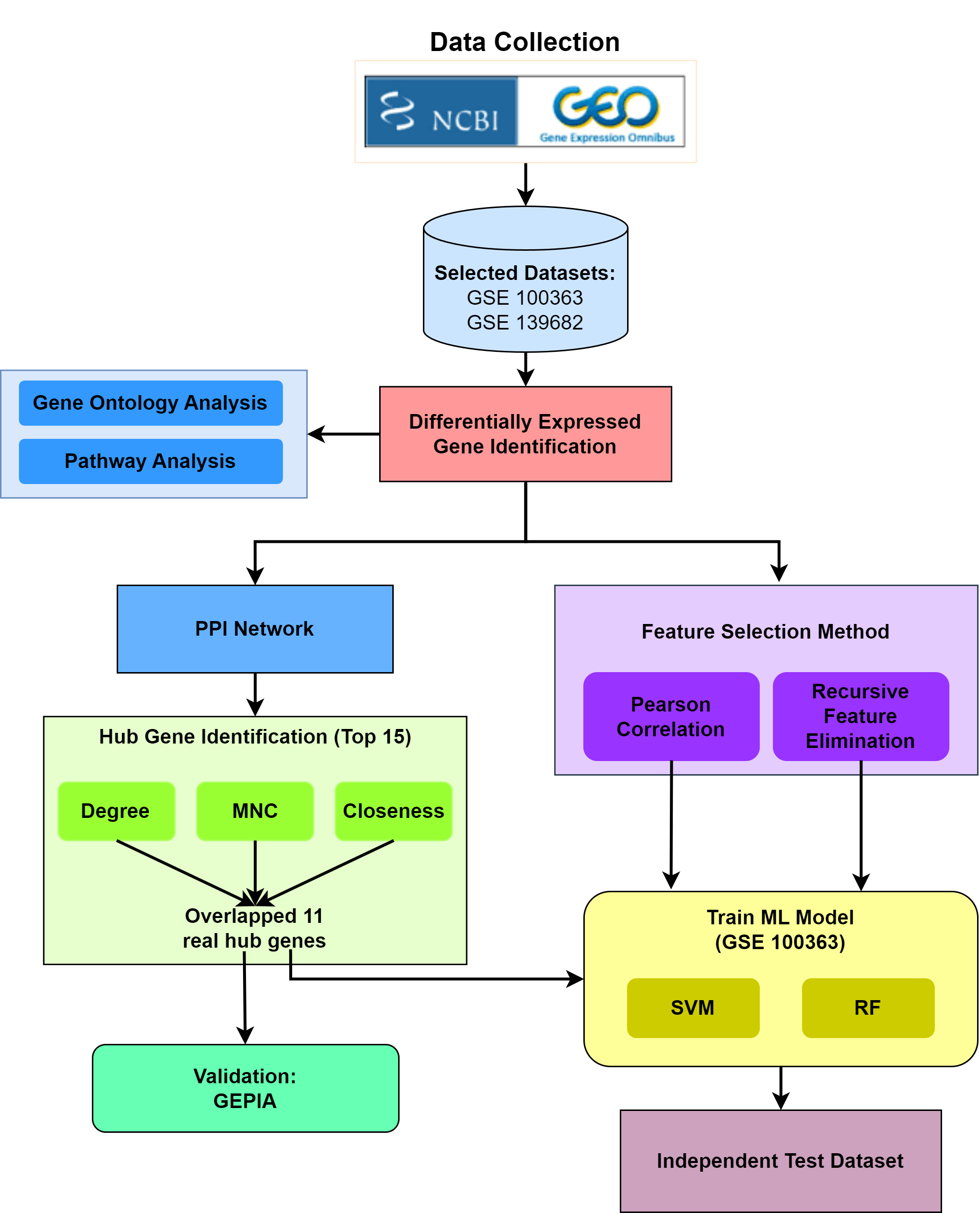}
    \captionsetup{labelfont=bf}
    \caption{\textbf{Flow diagram of proposed methodology:} Firstly, two microarray datasets (GSE100363, GSE139682) were downloaded from GEO. Secondly, differentially expressed genes (DEGs) were identified from those datasets. Next, the Gene Ontology analysis and Pathway analysis was performed with the identified DEGs to screen
significant GO terms and pathways. After that, the 
protein-protein interaction (PPI) network was constructed. 
 Subsequently, three ranking algorithms (Degree, MNC, Closeness 
Centrality) was employed to identify top 15 hub genes which, surprisingly, provided overlapped 11 real hub genes. In parallel, two feature selection methods (pearson correlation and recursive feature elimination)  was employed to further identify significant gene subsets.
Afterwards, the hub genes and significant genes subset were trained on GSE 100363 dataset
to develop machine learning model using SVM and RF algorithm. Finally, the model
was tested using independent GSE 139682 dataset to validate the biomarkers. Additionally, the real hub genes were validated using GEPIA database. }
    \label{fig:flow}
\end{figure}

This study is significant since it utilised bioinformatics and machine learning to analyse crucial genes for GBC and validate their diagnostic usefulness. This study is started by collecting 2 GBC-associated
microarray datasets from gene expression omnibus database. We employed bioinformatics to identify key DEGs in GBC from microarray datasets. DEGs are offered for further functional investigation and protein-protein interaction analysis. We also identified top 15 hub genes using
degree, maximum neighborhood component
(MNC), and closeness methods. The intersection of the 11 hub genes identified by the three methods was thought to hold the "real" hub genes. Additionally, significant genes were found using feature selection methods such as pearson correlation and the recursive feature elimination technique. Afterwards, the 11 real hub genes and significant genes, identified by feature selection method, are trained  on GSE 100363 dataset to develop machine learning model using support vector machine (SVM) and random forest (RF) algorithm. Finally, the model was tested using independent GSE 139682 dataset to validate the biomarkers. In addition, the 11 real hub genes were validated using GEPIA database.  Step by step process of this study is demonstrated in Figure \ref{fig:flow}.
\subsection{Research Gap}
Despite significant advancements in understanding gallbladder cancer (GBC), several gaps persist in previous research. While numerous studies have highlighted the essential roles of various biological processes in cancer spread and invasion, the detailed molecular mechanisms of GBC remain elusive. Specifically, prior research has identified key biological processes such as immune evasion \citep{tauriello2018tgfbeta}, epithelial-mesenchymal transition (EMT) \citep{wu2019epithelial} \citep{zheng2018elf3}, and cancer stem cells as critical in GBC progression, yet the precise pathways and interactions governing these processes are not fully understood. Current approaches have largely relied on traditional bioinformatics and high-throughput sequencing methods, such as microarrays and RNA sequencing, to identify potential diagnostic and prognostic biomarkers. There has been limited application of machine learning technologies to examine the predictive significance of differentially expressed genes (DEGs) in GBC. This lack of integration of machine learning methods may hinder the ability to achieve more reliable and reproducible outcomes. Moreover, while gene expression data have been extensively analyzed to identify relevant biomarkers for GBC, there remains a significant gap in understanding the enrichment pathways, Gene Ontology (GO) functions, and interaction networks of these DEGs. Previous studies have not thoroughly investigated these aspects, which are crucial for elucidating the underlying biological mechanisms and for developing effective therapeutic strategies. Addressing these gaps is essential for advancing the diagnosis, treatment, and prognosis of GBC, and for providing a more comprehensive understanding of this aggressive cancer.
\subsection{Research Question}
This study aims to identify and validate key biomarkers for gallbladder cancer (GBC) using a combination of bioinformatics and machine learning techniques. Specifically, it seeks to answer the following questions:
\begin{itemize}
    \item How can bioinformatics and machine learning approaches be used to identify biomarkers for gallbladder cancer (GBC) and which method produces optimal biomarkers for gallbladder cancer (GBC)?
    \item Which differentially expressed genes (DEGs) are significant in distinguishing between GBC and healthy samples?
    \item How can machine learning models be validated to accurately classify GBC samples based on these DEGs?
    \item What are the potential diagnostic and prognostic values of the identified hub genes and DEGs in GBC? 
\end{itemize}

These questions arise from the need to enhance the precision of diagnostic and prognostic biomarkers, which are crucial for the effective treatment and management of GBC. Additionally, these questions aim to bridge the gap between traditional clinical methods and modern computational techniques by leveraging the power of high-throughput gene expression data and advanced algorithms. Specifically, the study seeks to identify differentially expressed genes (DEGs) that can serve as reliable biomarkers for GBC, develop predictive models using these biomarkers, and validate the effectiveness of these models in distinguishing between healthy and cancerous samples.

\section{Methods}\label{sec2}

\subsection{Data Collection}
\begin{figure}[H]
    \centering
   \includegraphics[scale=0.3]{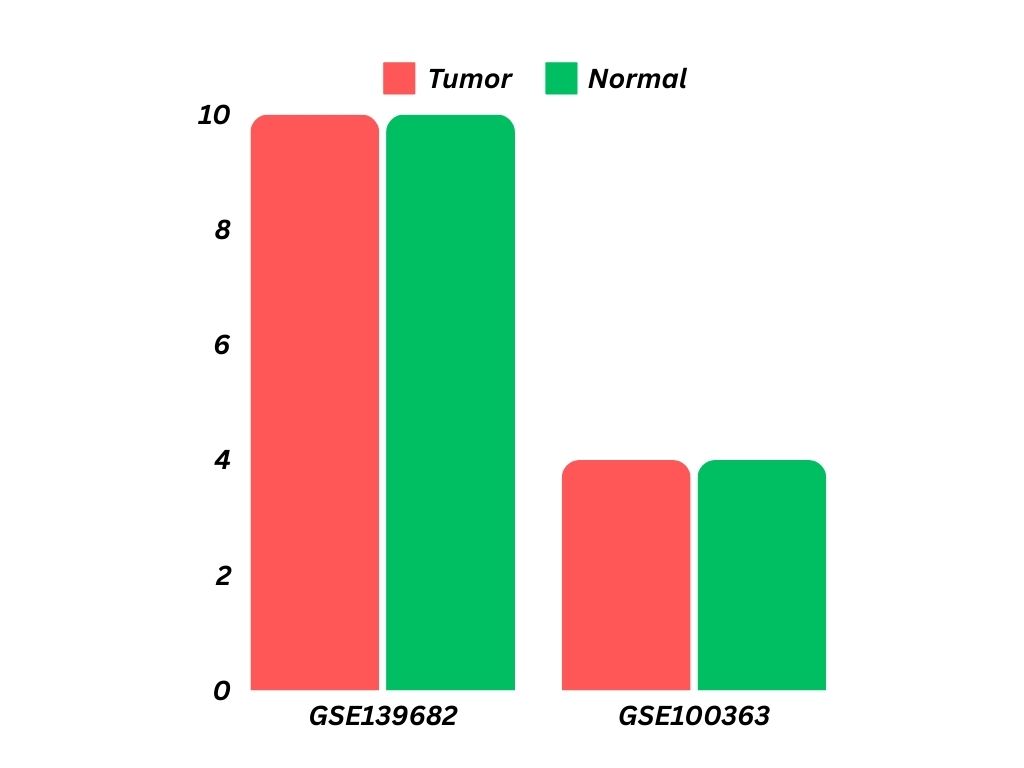}
    \caption{ Distribution of number of samples of both tumor and normal among selected datasets. The GSE139682 series comprises 20
samples: 10 gallbladder tumors and 10 normal samples. The GSE100363
series comprises 8 samples: 4 gallbladder tumours and 4 normal samples.  }
    \label{fig:data}
\end{figure}

Gene Expression Omnibus (GEO, https://www.ncbi.nlm.nih.gov/geo/) of National Center for Biotechnology Information (NCBI) \citep{barrett2012ncbi} is an accessible, high-throughput genome database that includes microarrays, chips, and gene expression data. The database, established in 2000, includes high-throughput gene expression data from research organisations worldwide. The repository provides access to published articles and associated gene expression detection data \citep{kong2023screening}. 
To find gene expression datasets for Gallbladder Cancer in the GEO database, we employed the search phrase "Gallbladder Cancer AND Homo sapiens" refined by "expression profling by array". We obtained a pair of datasets (GSE100363 and GSE139682) produced on the GPL20795 platform, HiSeq X Ten (Homo sapiens), containing gene expression data for those suffering from gall bladder tumours and normal people. 
The GSE100363 series comprises 8 samples: 4 gallbladder tumours and 4 normal samples. The GSE139682 series  comprises 20 samples:  10 gallbladder tumors and
10 normal samples. Figure \ref{fig:data} shows the distribution of tumour and normal sample counts across selected datasets..
\subsection{DEG Identification}
 Differentially expressed genes (DEGs) refer to genes whose expression levels change significantly between different experimental conditions, such as different tissues, developmental stages, or disease states. Upregulated genes show higher expression levels in one condition compared to another. Conversely,
 downregulated genes exhibit lower expression levels in one condition compared to another. The upregulation of genes can indicate activation of specific biological processes, response to environmental stimuli, or involvement in disease pathways. On the other hand, the downregulation of genes may indicate suppression of certain biological processes, adaptation to environmental changes, or inhibition of disease pathways.  
 
 GEO2R (http://www.ncbi.nlm.nih.gov/geo/geo2r), a data analysis programme included within the GEO database, allows for visual statistical analysis of gene expression profiles. It examines multiple GEO series datasets to discover DEGs under experimental settings \citep{xu2016identification}. We utilised GEO2R  to identify DEGs in both GBC tumour and normal samples. Utilising the GEO database, we acquired the gene $log2 fold change (log2 FC)$ value. The ratio of the gene expression values for cancer and normal is expressed as a fold change of the gene. The upregulated and downregulated genes were determined by the positive and negative values of 
the $log2FC$ values. Adjusted $ P-value < 0.05 $ and $|log2 fold change (log2 FC) | > 1 $ for overexpression along with $ |log2 fold change (log2 FC) | < -1 $ for downexpression are employed to determine the statistically significant effect of genes. DEGs were visualized using a volcano map. Venn diagrams were created using the Venny online tool (https://bioinfogp.cnb.csic.es/tools/venny/) \citep{oliveros2007venny}.

\subsection{Gene Ontology and Pathway Enrichment Analysis of DEGs}

Gene Ontology (GO) study yields broad biological research results for a particular gene or gene set by utilising the concepts of molecular functions (MFs), biological processes (BPs), and cellular components (CCs). The study of GO has become an essential component of investigations connected to system biology in recent times.
Pathway enrichment analysis is another tool that helps investigate the biological relationships between gene sets derived from extensive genome-scale investigation \citep{reimand2019pathway}. In this work, the Gene Ontology database \citep{ashburner2000gene} was utilised to investigate the GO terms related with DEGs, and the REACTOME \citep{croft2010reactome} databases were utilised to do pathway analysis. 
An online biological information directory called the Database for Annotation, Visualisation, and Integrated Discovery (DAVID, http://david.abcc.ncifcrf.gov/) combines biological data and analysis tools to offer a complete collection of annotation information for functional genes and proteins \citep{huang2007david}. It offers a way for users to extract biological data. The DAVID online database was used to do biological studies and examine how DEGs work \citep{ashburner2000gene}. 

\subsection{Protein–Protein Interaction (PPI) network 
construction}
A public database called the Search Tool for the Retrieval of Interacting Genes/Proteins (STRING; http://string-db.org) is employed to anticipate the PPI networks\citep{franceschini2012string}.
Data on over 5000 species, 20 million proteins, and 3 billion interactions can be found in the STRING database. Such protein interactions encompass both co-expression correlations and direct physical interactions. To gain more knowledge of the intricate regulatory networks seen in organisms, known protein-protein connections can be discovered using the STRING database. According to earlier research, examining the functional relationships between proteins can reveal fresh information on the origin or progression of disorders\citep{smoot2011cytoscape}. Applying the STRING database, the PPI network of DEGs was created for this study.

\subsection{Hub gene Identification}
Hub genes are often utilised to focus on the subset of DEGs that would most effectively separate the ill samples from the control group. In turn, we visualised the constructed PPI network using the Cytoscape software \citep{kohl2011cytoscape} (https://cytoscape.org/), and we used the Cytoscape CytoHubba plugin to identify the hub genes in the PPI network using a variety of ranking techniques.

Various methods for node ranking, including local and global approaches, are offered by CytoHubba \citep{chin2014cytohubba}. The global approach looks at the node's interaction with the entire network, whereas the local ranking technique looks at the node's interaction with its immediate peers. The hub genes were identified using three ranking algorithms: a global ranking method (Closeness Centrality), a pair of local ranking algorithms (Degree and Maximum Neighbourhood Component (MNC)). The quantity of nodes that surround a node v is its degree. Maximum connected component (MNC) is the size of the neighbourhood N(v), which is the set of nodes that are next to v but do not contain v. Last but not least, closeness centrality, which is determined by averaging the length of the shortest path connecting a node to every other node in the network, shows how close a node is to every other node in the network. In order to create machine learning models, we finally determined the top 15 genes from each ranking approach.

\subsection{Identification of significant DEGs using feature selection methods}
We used feature selection techniques, such as Pearson Correlation and Recursive Feature Elimination (RFE), which select the best features in high-dimensional data, to further find the important DEGs that best distinguish the diseased samples from the healthy controls. In a nutshell, RFE ranks the features according to their significance and returns the top-n features after eliminating the less significant ones, where n is the number of features that the user entered. In order to utilise it, the $estimator$ parameter needs to be initialised by indicating the algorithm to be used and the $n\_features\_to\_select$ parameter by indicating the number of features to be selected. Following configuration, the model must be fitted to a training dataset using the $fit()$ method in order to select the features. In our research, we supplied the sample matrix containing all of the features (146 DEGs) into the RFE model, setting $n\_features\_to\_select$ to 20 and using the SVM method as a $estimator$. Furthermore, Pearson correlation demonstrates the linear relationship between two variables.Strongly correlated features have a more linear dependence and so have a similar effect on the dependent variable. Two qualities may be eliminated if there is a substantial correlation between them. In our research, we entered the sample matrix including all 146 DEGs of features into a Pearson correlation model, compared the feature correlations, and removed one of the two features with a correlation higher than 0.9.

\subsection{Developing ML models on GSE100363 dataset}
For distinction among GBC tumours and healthy samples, we created two machine learning models i.e., Random Forest (RF) and Support Vector Machine (SVM) classification models. The applications of SVM, a supervised learning method, is in regression and classification. For classification purposes, the SVM generates hyperplanes that maximise the separation between classes. On the other hand, an ensemble classifier called Random Forest \cite{breiman2001random} \citep{khatun2023cancer} consists of many decision trees and produces a class that is the mode of the output of each tree individually.  In every classification tree, a certain amount of votes are assigned to each class. Out of all the trees, the algorithm chooses the category with the highest number of votes.

Moreover, we developed a space of searches for parameter optimisation for each machine learning model in order to determine the optimal set of attributes. As a result, for hyperparameter optimisation in RF and SVM, we employed a grid search technique after randomised search. Regarding the randomised search, we generated a grid of hyperparameters, and we used random hyperparameter combinations to train and test our models.Subsequently, the optimal parameter combinations are determined by identifying the best parameters using randomised search.
Table \ref{tab:parameter} summarizes the tested parameters and selected parameters (highlighted in bold).
We employed 4-fold-cross-validation (CV) to assess such models. The StratifedKFold method is used to divide the data into four segments, assuring that each subgroup has an equal proportion of positive and negative observations. Choosing one of the subsets for testing and using all the other subsets for training allows the process to be performed four times. We computed the accuracy score for every fold and utilised that result for calculating the mean accuracy.

\begin{table}[h]
\centering
\begin{tabular}{|p{3cm}|p{4cm}|p{4cm}|}
\hline
\textbf{Model }               & \textbf{Hyperparameters}       & \textbf{Search Space}                  \\ \hline \hline
\multirow{3}{*}{SVM} & C & {[}\textbf{0.1}, 1, 10, 100{]} \\
 & Kernel &{[}\textbf{'linear'},'rbf', 'poly'{]} \\ 
 & Gamma & {[}\textbf{'scale'}, 'auto', 0.1, 1{]} \\ \hline
\multirow{4}{*}{RF}  & 'n\_estimators'       & {[}\textbf{50}, \textbf{100}, 200{]}            \\  
& 'max\_depth'          & {[}\textbf{None}, 10, 20, 30{]}        \\ 
 & 'min\_samples\_split' & {[}\textbf{2}, 5, 10{]}                \\ 
 & 'min\_samples\_leaf'  & {[}\textbf{1}, 2, 4{]}                 \\ \hline
\end{tabular}
\caption{Search space parameters for RF and SVM model optimisation. The best parameter values are highlighted in bold.}
\label{tab:parameter}
\end{table}

 \subsection{Evaluation Tools}
 Our proposed methodology was assessed using ROC curve performance matrices and accuracy score \citep{fakoor2013using}. The accuracy score is a measure of the correctness of the model's predictions. It quantifies the percentage of correct predictions made by the model out of the total predictions made. The following formula is used to determine the accuracy score:

 \begin{equation}
 Accuracy\; score = \frac{Number\; of\; Correct\; Predictions}{Total\; Number\; of\; Predictions} \times 100\%
\end{equation}

A graphical representation known as the Receiver Operating Characteristic (ROC) curve is used to assess how well a binary classification model performs across various thresholds. For different threshold settings, it plots the actual positive rate (sensitivity) against the false positive rate (specificity). Specifically, lower values on the plot's x-axis indicate lower false positives and higher true negatives, whereas higher values on the figure's y-axis indicate higher true positives and lower false negatives. The area under the ROC curve (AUC-ROC) is commonly used as a summary metric to evaluate the overall performance of the classifier, with values ranging from 0 to 1. A higher AUC-ROC indicates better discrimination capacity of the model.

\subsection{Validation of biomarkers gene expression}
Using GEPIA2 (Gene Expression Profiling Interactive Analysis, http://gepia2.cancer-pku.cn/), a database of information obtained from the UCSC Xena server that contains 9736 tumour samples and 8587 normal samples, the expression levels of biomarkers in GBC and normal cases were confirmed. P-values less than 0.05 denoted statistically significant variations \citep{tang2017gepia}.

\section{Results}\label{sec3}

\subsection{DEG Identification }
\begin{figure}[]
    \centering    
    \subfigure[]{\includegraphics[scale=.20]{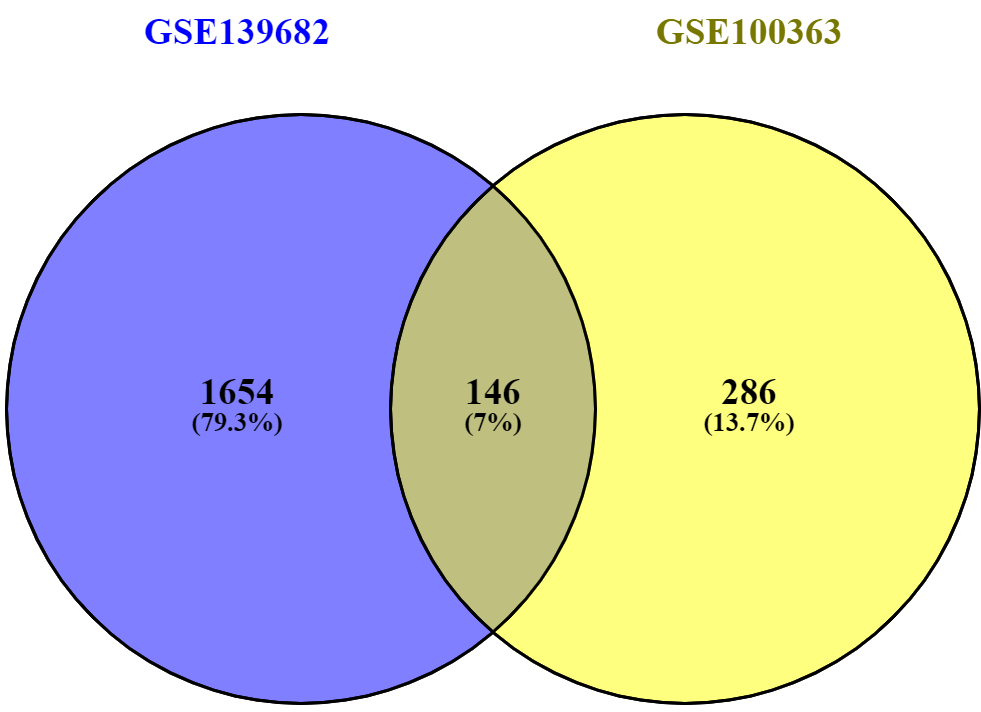}}
    \subfigure[]{\includegraphics[scale=.20]{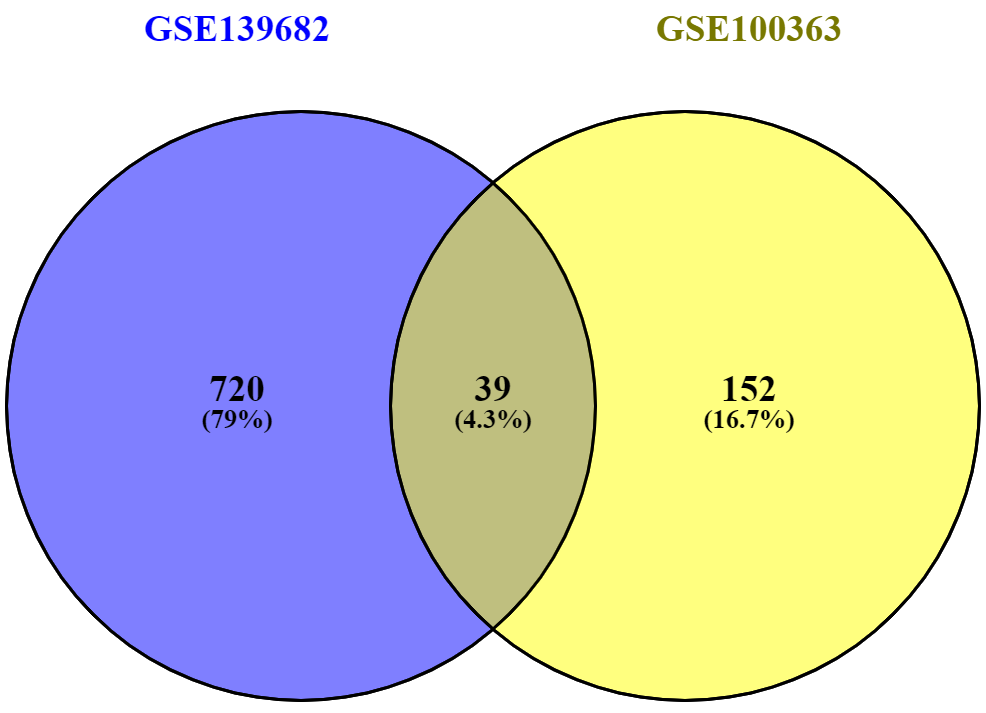}}
    \subfigure[]{\includegraphics[scale=.20]{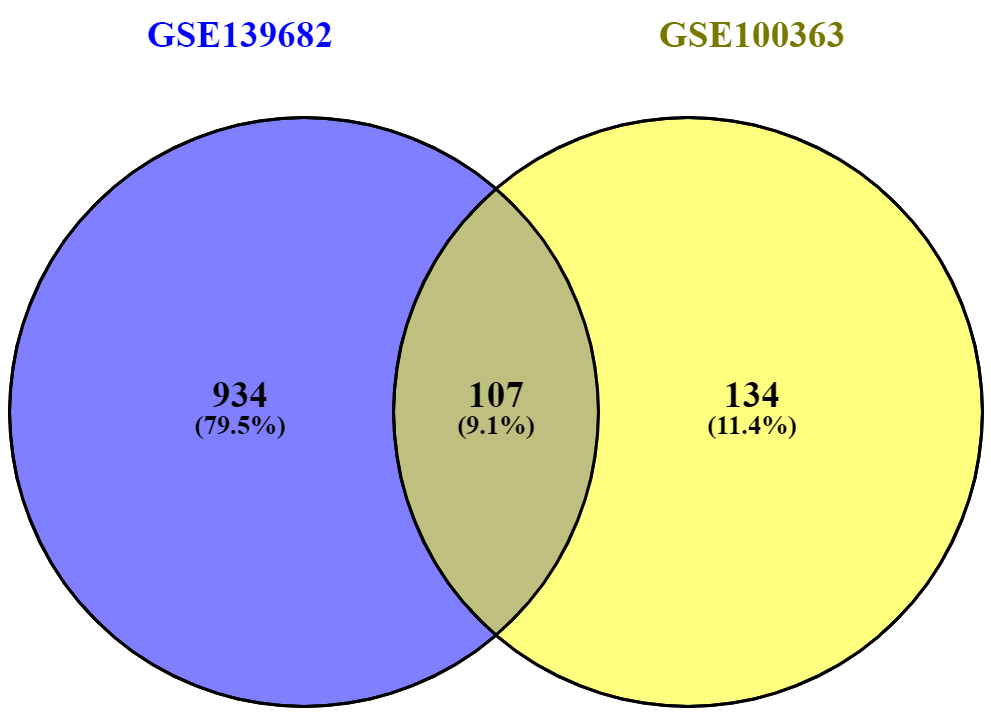}}
   \captionsetup{labelfont=bf}
    \caption{Venn intersection diagrams of the DEGs of the two datasets: (a) represents the common DEGs, and (b) represents the
common upregulated genes.(c) represents the
common downregulated genes.}
    \label{fig:conf1}
\end{figure}

\begin{figure}[]
    \centering
    \subfigure[GSE100363]{\includegraphics[scale=.4]{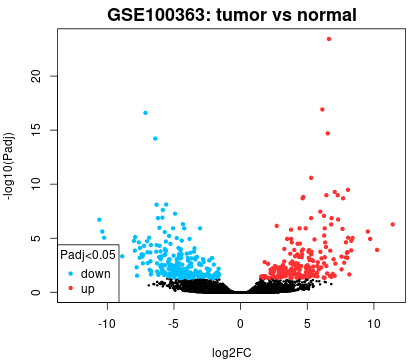}
 }
    \subfigure[GSE139682]{\includegraphics[scale=.4]{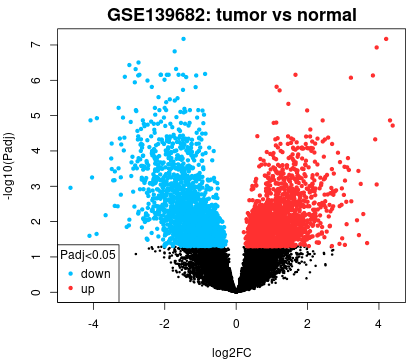}}
   
    \caption{Volcano plots of differentially expressed genes. Blue data points indicates down-regulated 
genes and red data points indicates up-regulated genes.  The $|logFC| > 1 $ for overexpression and $|logFC| <-1  $ for downexpression was applied to set up differences. Black dots indicated genes that were not 
differentially expressed.}
    \label{fig:volcano}
\end{figure}

Initially from the GSE139682 and GSE100363 datasets, respectively, a total of 22829 and 22031 DEGs were discovered. Following execution of the revised $P$ value requirement and minimal $log2FC$, 1800, 432 DEGs are found in accordance (Table \ref{tab:DEG}). In the end, the common differential expression analysis between the tissues from gallbladder tumours and normal tissues produced 146 common differentially expressed genes among which the expression of 39 genes 
was found to be significantly up-regulated and 107 genes 
were down-regulated in GSE100363 and GSE139682 (Figure: \ref{fig:conf1}). The volcano plot of different expression were shown in (Figure:\ref{fig:volcano}).

\begin{table}[h]
\resizebox{\textwidth}{!}{
\begin{tabular}
{c|c|ccc|}

\hline
\multicolumn{1}{|l|}{ \multirow{2}{*}{\textbf{Accession Number}}}                  & \textbf{\underline{Before log2FC and adjusted p value filtration}} & \multicolumn{3}{c|}{\textbf{\underline{After log2FC and adjusted p value filtration}}}                                           \\   
\multicolumn{1}{|c|}{} & \textbf{Total DEGS}                           & \multicolumn{1}{c|}{\textbf{Total DEGS}} & \multicolumn{1}{c|}{\textbf{Upregulated DEGS}} & \textbf{Downregulated DEGs} \\ \hline
\multicolumn{1}{|c|}{GSE139682}        & 22829                                         & \multicolumn{1}{c|}{1800}                & \multicolumn{1}{c|}{759}                       & 1041               \\ \hline
\multicolumn{1}{|c|}{GSE100363}        & 22031                                         & \multicolumn{1}{c|}{432}                 & \multicolumn{1}{c|}{191}                       & 241                \\ \hline
\multicolumn{1}{|c|}{Overlapped}       & 20952                                         & \multicolumn{1}{c|}{146}                 & \multicolumn{1}{c|}{39}                        & 107                \\ \hline

\end{tabular}
}

\caption{Dataset analysis details before filtration and after
log2FC and adjusted p value  filtration.}
\label{tab:DEG}
\end{table}

\begin{table}[]
\resizebox{\textwidth}{!}{
\begin{tabular}{llll}
\hline
\textbf{Category} & \textbf{GO ID}                          & \textbf{\%} & \textbf{P value} \\ \hline
BP& Epidermis Development                    & 7.692308    & 0.006651         \\ \hline
BP       & Keratinization                           & 7.692308    & 0.007428         \\ \hline
BP       & Endodermal Cell Differentiation          & 5.128205    & 0.043789         \\ \hline
BP      & Cell Adhesion                            & 10.25641    & 0.046459         \\ \hline
BP     & Establishment of Skin Barrier            & 5.128205    & 0.04655          \\ \hline
BP       & Intermediate Filament Organization       & 5.128205    & 0.093636         \\ \hline
CC         & Ciliary   Membrane                       & 5.128205    & 0.085086         \\ \hline
CC       & Cornified Envelope                       & 5.128205    & 0.08375          \\ \hline
CC     & Intermediate Filament   Cytoskeleton     & 5.128205    & 0.082411         \\ \hline
CC       & Integral Component of Membrane           & 33.33333    & 0.068537         \\ \hline
CC    & Hippocampal Mossy Fiber to CA3   Synapse & 5.128205    & 0.060735         \\ \hline
CC            & Integral Component of Plasma   Membrane  & 15.38462    & 0.054364         \\ \hline
CC   & Cytoskeleton                             & 10.25641    & 0.044753         \\ \hline
CC         & Occluding Junction                       & 5.128205    & 0.037162         \\ \hline
CC        & Apical Plasma Membrane                   & 10.25641    & 0.019177         \\ \hline
CC         & Intermediate Filament                    & 7.692308    & 0.017341         \\ \hline
CC              & Plasma Membrane                          & 38.46154    & 0.011103         \\ \hline
MF        & Structural Molecule Activity             & 10.25641    & 0.003175         \\ \hline
REACTOME   & Signaling by Receptor Tyrosine Kinases     & 10.25641026 & 0.051975399 \\ \hline
REACTOME    & Formation of the Cornified Envelope      & 7.692307692 & 0.018876409  \\ \hline
 REACTOME  & Keratinization & 7.692307692  & 0.047712584 \\ \hline
 REACTOME   & Extracellular Matrix Organization   & 7.692307692  & 0.085736649  \\ \hline
 REACTOME              & Anchoring Fibril Formation            & 5.128205128 & 0.024375819 \\ \hline
  REACTOME              & LDL Clearance            & 5.128205128  & 0.030780624 \\ \hline
 REACTOME              & Laminin Interactions            & 5.128205128  & 0.048189704 \\ \hline
 REACTOME              & Plasma Lipoprotein Clearance            & 5.128205128 & 0.059113989 \\ \hline
 REACTOME              & Non-integrin Membrane-ECM Interactions            & 5.128205128 & 0.092681663 \\ \hline
 REACTOME              & Assembly of Collagen Fibrils and other Multimeric Structures            & 5.128205128 & 0.095676523 \\ \hline
         
\end{tabular}
    }
\caption{Gene Ontolgy analysis of upregulated DEGs using DAVID web tool.}
\label{tab:GO-up}

\end{table}

\begin{table}[]
\resizebox{\textwidth}{!}{
\begin{tabular}{p{2cm}p{9cm}p{2cm}p{2cm}}
\hline
\textbf{Category} & \textbf{GO ID}                                                                                                       & \textbf{\%} & \textbf{P value} \\ \hline
BP         & Cell   Differentiation                                                                                                 & 12.38095    & 1.10E-04         \\ \hline
BP          & Nervous System Development                                                                                             & 10.47619    & 3.16E-05         \\ \hline
BP        & Cell Adhesion                                                                                                          & 8.571429    & 0.005604         \\ \hline
BP          & Cell-Cell Adhesion                                                                                                     & 6.666667    & 3.51E-04         \\ \hline
BP          & Extracellular Matrix   Organization                                                                                    & 5.714286    & 0.002013         \\ \hline

CC        & Plasma   Membrane                                                                                                      & 40.95238    & 1.40E-04         \\ \hline
CC           & Integral Component of Membrane                                                                                         & 37.14286    & 0.003353         \\ \hline
CC             & Integral Component of Plasma   Membrane                                                                                & 19.04762    & 3.49E-05         \\ \hline
CC          & Extracellular Space                                                                                                    & 19.04762    & 0.001877         \\ \hline
CC          & Extracellular Region                                                                                                   & 17.14286    & 0.02544          \\ \hline
CC        & Cell Surface                                                                                                           & 10.47619    & 0.001011         \\ \hline
CC              & Axon                                                                                                                   & 8.571429    & 2.99E-04         \\ \hline
CC            & Dendrite                                                                                                               & 7.619048    & 0.005228         \\ \hline
CC             & External Side of Plasma   Membrane                                                                                     & 7.619048    & 0.00729          \\ \hline

MF        & Calcium   ION Binding                                                                                                  & 8.571429    & 0.027692         \\ \hline
MF            & Heparin Binding                                                                                                        & 7.619048    & 2.39E-05         \\ \hline
MF           & Actin Binding                                                                                                          & 4.761905    & 0.081892         \\ \hline
MF        & Structural Molecule Activity                                                                                           & 3.809524    & 0.066818         \\ \hline
MF            & Carbohydrate Binding                                                                                                   & 3.809524    & 0.073484         \\ \hline
MF        & Signaling Receptor Activity                                                                                            & 3.809524    & 0.094202         \\ \hline

REACTOME & Extracellular Matrix   Organization            & 8.571429 & 1.14E-04 \\ \hline
REACTOME & Collagen Biosynthesis and Modifying enzymes    & 3.809524 & 0.004569 \\ \hline
REACTOME & Cell Surface Interactions at the Vascular Wall & 4.761905 & 0.004872 \\ \hline
REACTOME & Collagen Formation                             & 3.809524 & 0.010334 \\ \hline
REACTOME & Muscle Contraction                             & 4.761905 & 0.019309 \\ \hline
REACTOME & NGF-independant TRKA Activation                & 1.904762 & 0.024852 \\ \hline
REACTOME & Cardiac Conduction                             & 3.809524 & 0.028455 \\ \hline
REACTOME & Activation of TRKA Receptors                   & 1.904762 & 0.029749 \\ \hline
REACTOME & Degradation of the Extracellular Matrix        & 3.809524 & 0.03306  \\ \hline
REACTOME & Ligand-receptor Interactions                   & 1.904762 & 0.039471 \\ \hline
REACTOME & Neuronal System                                & 5.714286 & 0.054045 \\ \hline
REACTOME & Integrin Cell Surface Interactions             & 2.857143 & 0.067809 \\ \hline
REACTOME & Signal Transduction                            & 18.09524 & 0.079522 \\ \hline
REACTOME & Activation of SMO                              & 1.904762 & 0.086664 \\ \hline
REACTOME & Hemostasis                                     & 6.666667 & 0.089655 \\ \hline
REACTOME & Scavenging by Class A Receptors                & 1.904762 & 0.091256 \\ \hline

\end{tabular}
}
\caption{Gene Ontolgy analysis of downregulated DEGs using DAVID web tool.}
\label{tab:GO-down}
\end{table}

\begin{figure}[]
    \centering
   \includegraphics[scale=0.8]{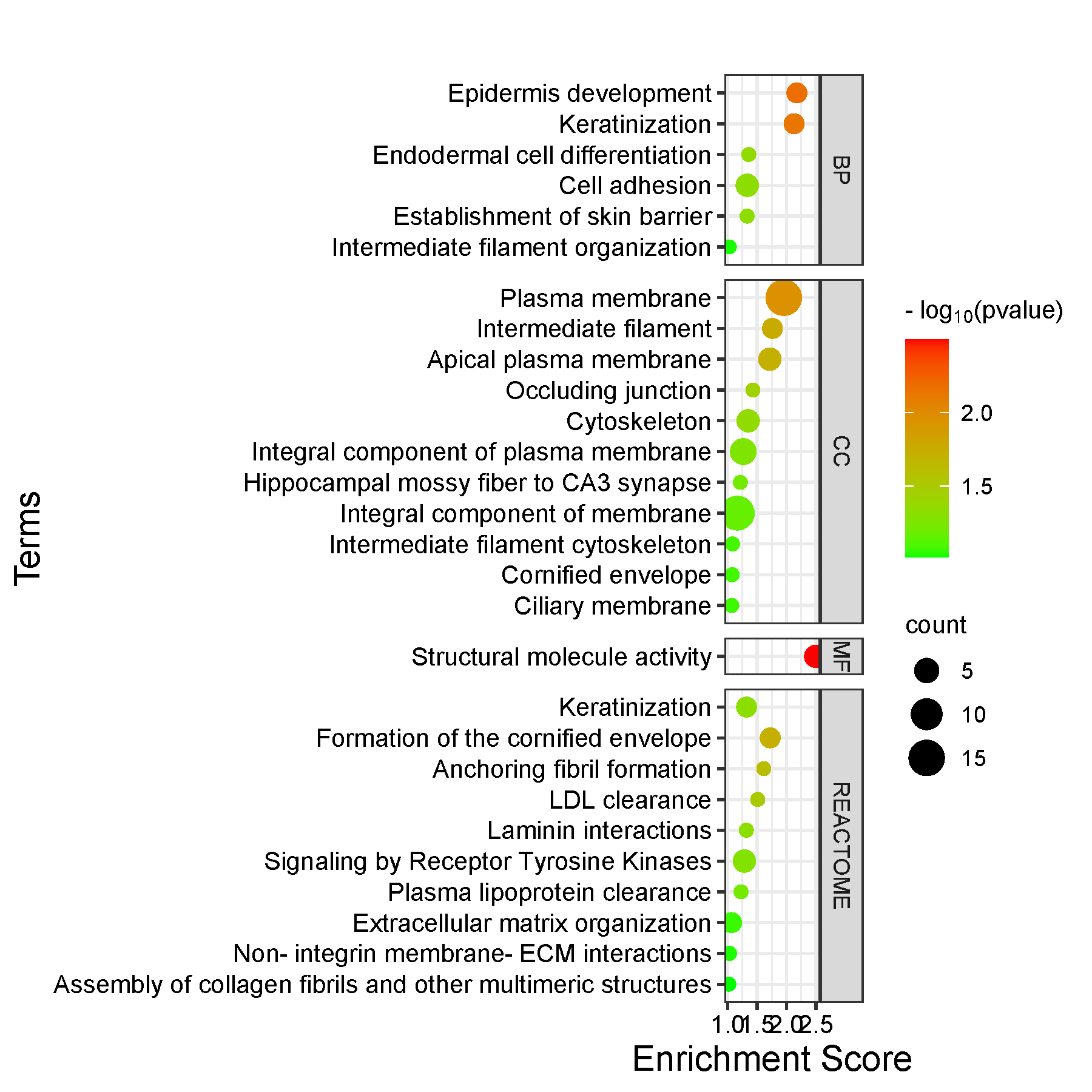}
    \caption{ \textbf{Functional enrichment analyses:} GO terms and REACTOME pathways for upregulated DEGs in this study. \textbf{BP:} Upregulated DEGs enriched in biological Process. \textbf{CC:} Upregulated DEGs enriched in cellular component. \textbf{MF:} Upregulated DEGs enriched in molecular function. The size of the bubble indicates the enrichment 
score; colors indicate enrichment signifcane. 
}
    \label{fig:gpu}
\end{figure}

\subsection{Gene Ontology and Pathway Enrichment Analysis of DEGs} 

\begin{figure}[]
    \centering
   \includegraphics[scale=0.8]{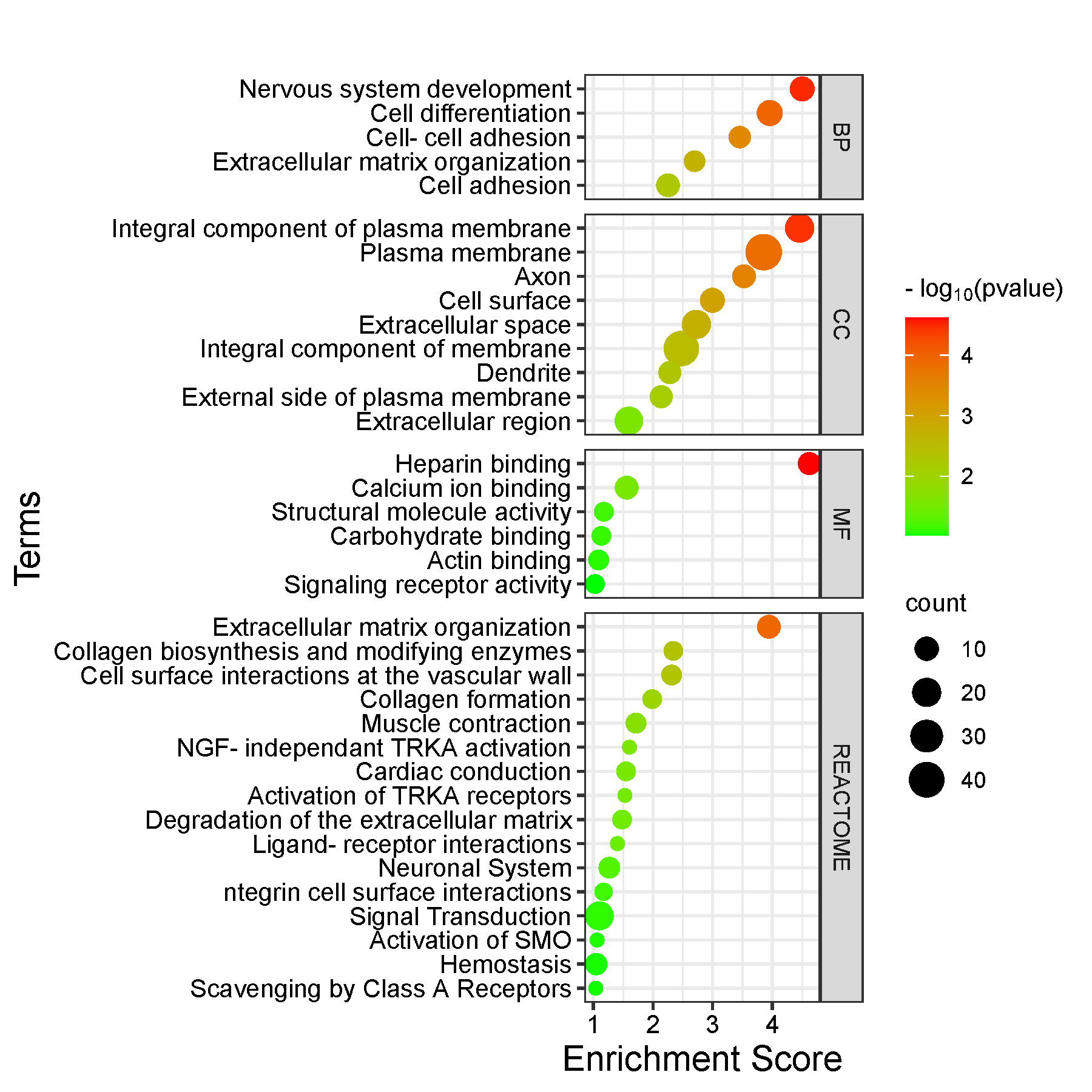}
    \caption{Functional enrichment analyses: GO terms for downregulated differentially expressed genes 
(DEGs) in this study and REACTOME pathways for DEGs. The size of the bubble indicates the enrichment 
score; colors indicate enrichment signifcane}
    \label{fig:gpd}
\end{figure}

 Using the DAVID online tool, GO functional and pathway enrichment analyses were carried out on the 146 common DEGs. 
To learn more about the function of the DEGs that were found, functional analysis was used. Significantly enriched GO terms and pathways of the discovered DEGs are revealed by the functional analysis.
Based on the GO study, it can be inferred that the overexpressed DEGs are primarily linked to biological processes such as "Cell adhesion", "Epidermis development", and "Keratinization"; cellular components such as the "plasma membrane" and "integral membrane"; and molecular functions such as "structural molecule activity" (Table \ref{tab:GO-up}). Likewise, the investigation reveals that the downexpressed DEGs are primarily linked to "cell differentiation", "nervous system development", "cell adhesion" for biological process, "the plasma membrane", "integral component of membrane" for cellular component, "calcium ion binding", and "heparin binding" for molecular function. (Table \ref{tab:GO-down}). To further explore the pathways that were found to be enriched in DEGs, we performed a REACTOME pathway enrichment analysis next. The REACTOME pathway analysis revealed that the pathways involved in overexpression of DEGs were primarily enriched in "Signalling by Receptor Tyrosine Kinases" , "Formation of the Cornified Envelope" , "Keratinization" , and "Extracellular Matrix Organisation" pathway (Table \ref{tab:GO-up}). Furthermore, "Signal Transduction" , "Extracellular Matrix Organisation" , and "Hemostasis" pathways were the key ones where downexpressed DEGs were enriched (Table \ref{tab:GO-down}).Using an enrichment dot bubble diagram made with the bioinformatics website and GO-enriched BP/MF/CC and REACTOME, the GO keywords and pathways for both overexpressed DEGs (Figure \ref{fig:gpu}) and downexpressed DEGs (Figure \ref{fig:gpd}) were visualised.

\subsection{PPI Construction and Screening for Hub Genes}
\begin{figure}
   
    \centering
    \includegraphics[scale=0.25]{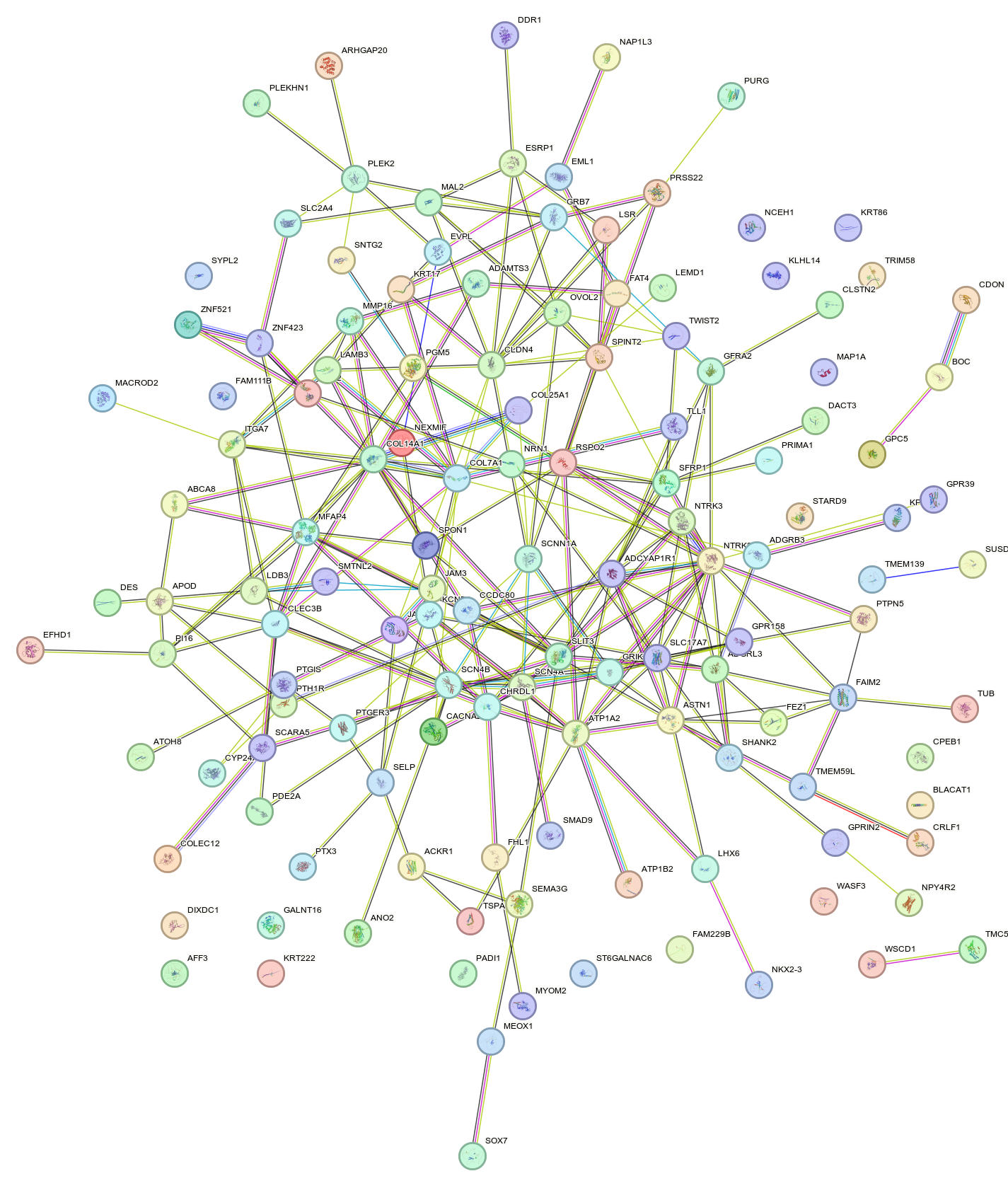}
    \caption{PPI network of 146 promising target genes in gall bladder cancer based on the string website.}
    \label{fig:diagram}
\end{figure}

\begin{figure}[]
    \centering
    \subfigure[Degree]{\includegraphics[scale=.15]{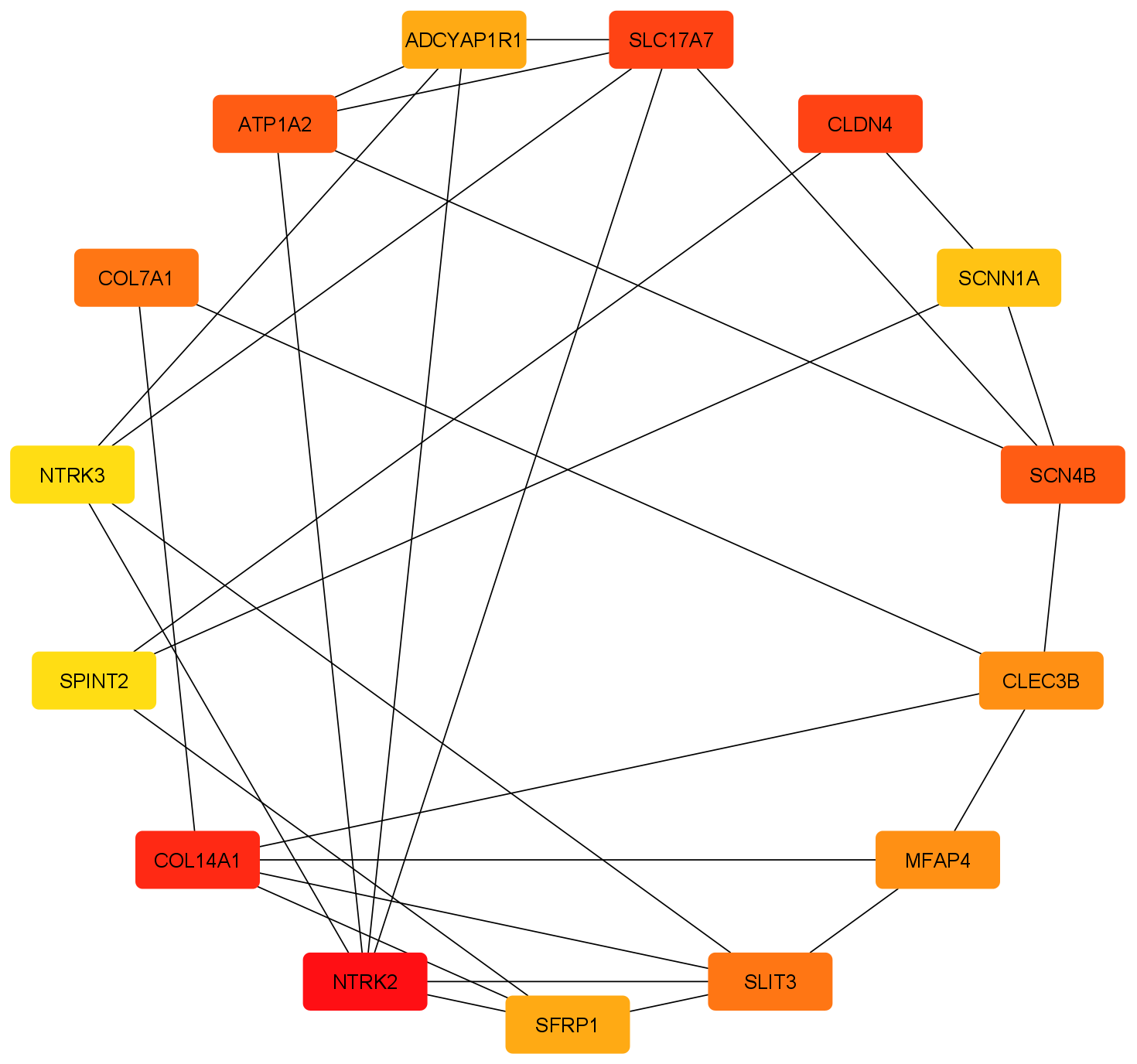}}
    \subfigure[Closeness]{\includegraphics[scale=.15]{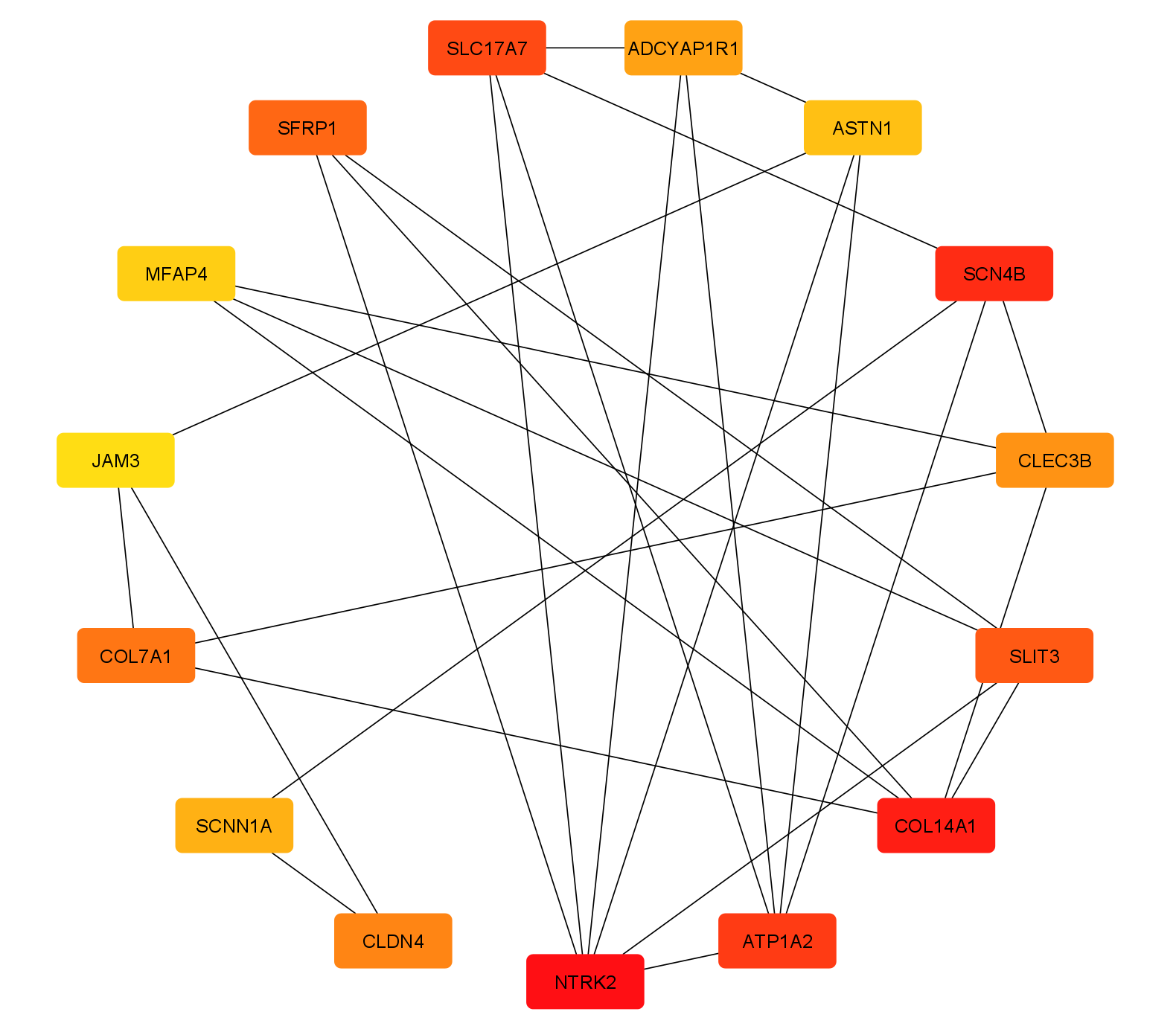}}
    \subfigure[MNC]{\includegraphics[scale=.15]{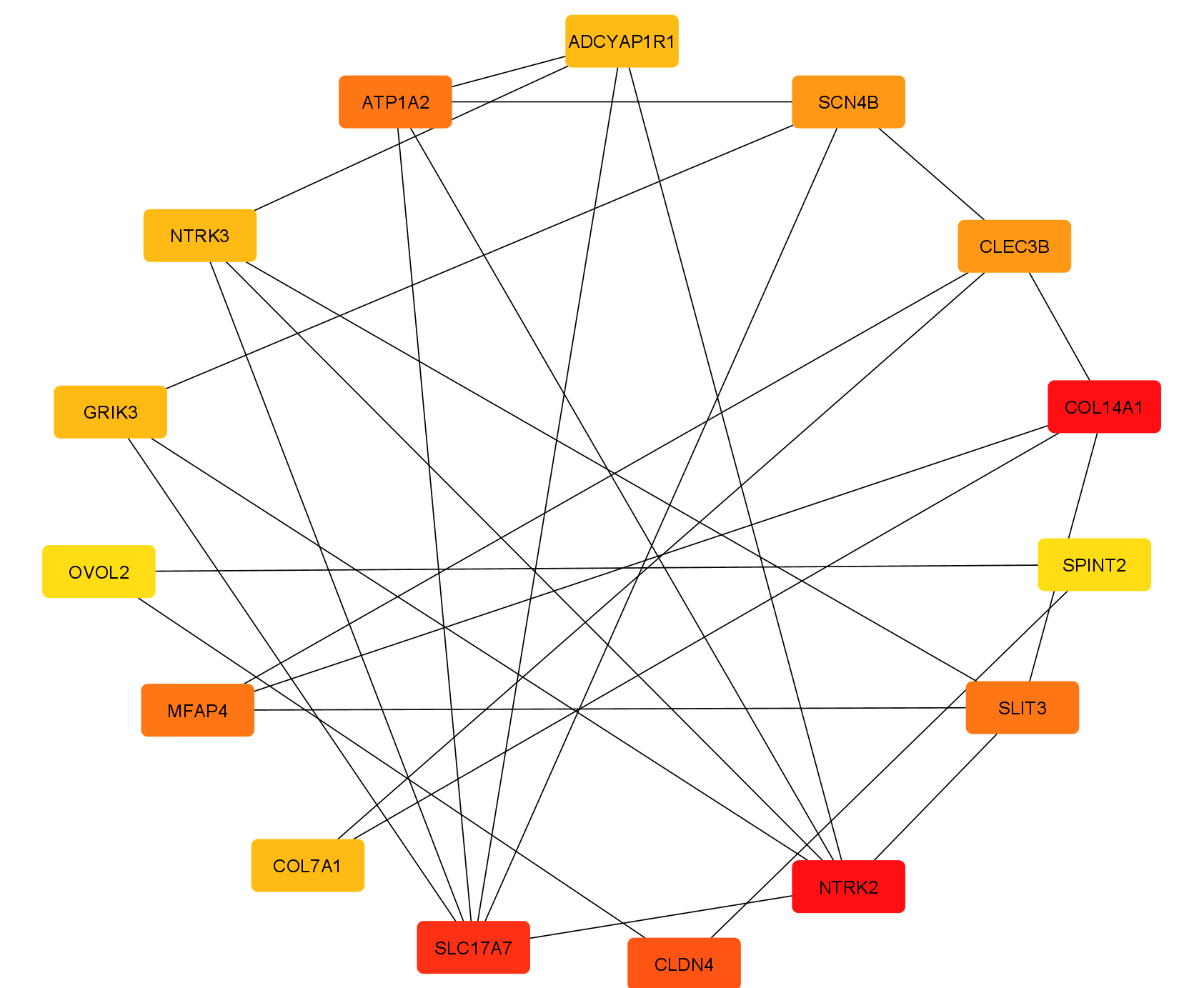}}
    \subfigure[Venn Intersection Diagram]{\includegraphics[scale=.4]{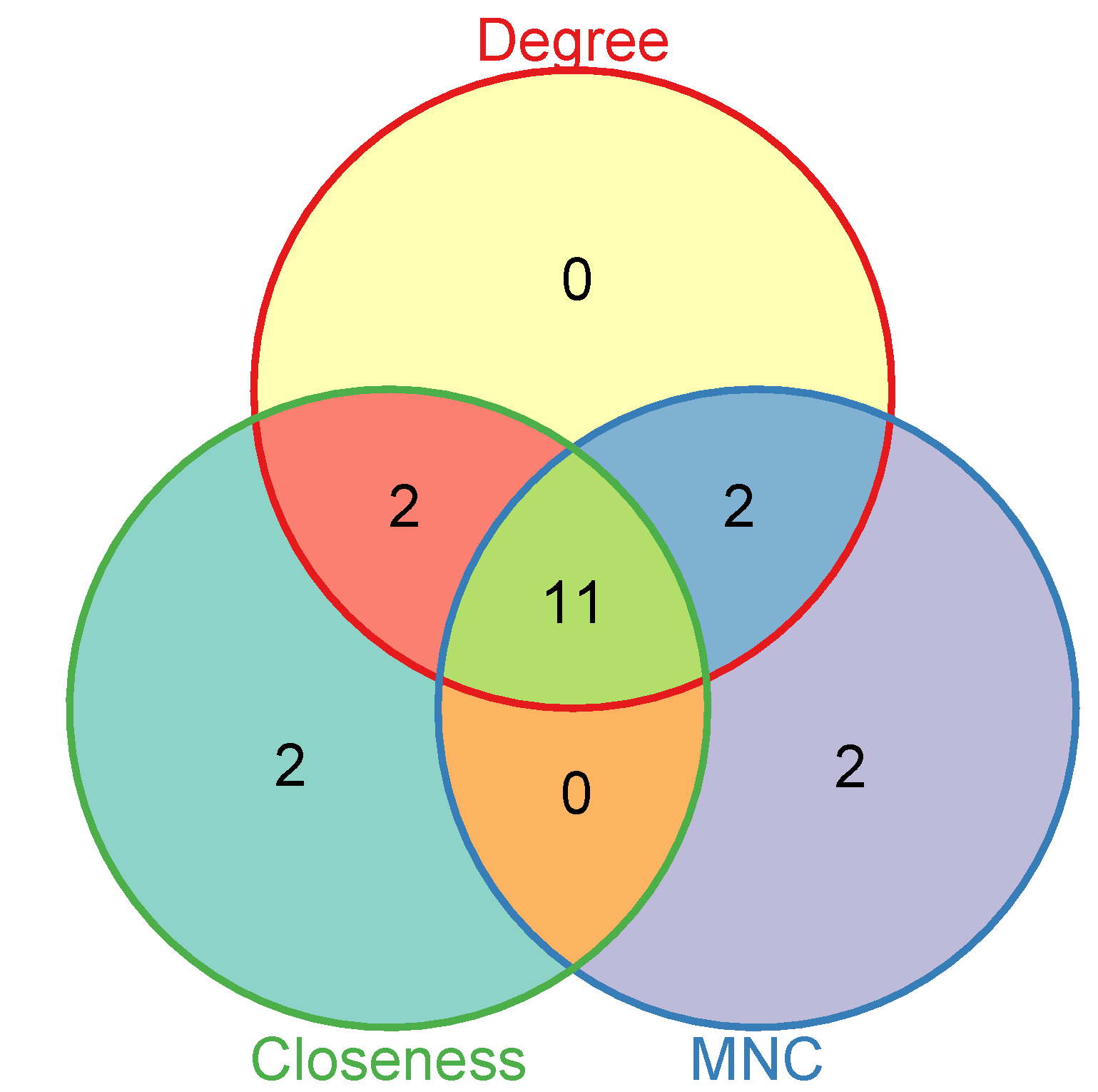}}
    \caption{(a-c) Hub genes identification with the CytoHubba plugin in Cytoscape software. Three different metrics were used: DEGREE, MNC, closeness. (d) A Venn diagram used to identify 11 hub genes in gall bladder cancer.}
    \label{fig:conf}
\end{figure}

Utilising Cytoscape and the STRING online database, we constructed a DEG PPI network to investigate the protein networks linked to identified genes (Figure \ref{fig:diagram}).
Next, employing the cytoHubba plugin, the PPI network was loaded into the Cytoscape software to visualise and identify the hub genes. We subsequently acquired the top-15 hub genes for each of the three topological ranking algorithms: Closeness Centrality, Maximum Neighbourhood Component (MNC), and Degree (Figure \ref{fig:conf}). Amazingly, we discovered that just 11 of the hub genes were identified by all three ranking methods (Figure \ref{fig:conf}, Table \ref{tab:hub}). The 11 "real" hub genes were found at the intersection of the hub genes produced by the three methods. 

\begin{table}[h]
\resizebox{\textwidth}{!}{
\begin{tabular}{|
>{\columncolor[HTML]{FFFC9E}}l |l|}
\hline
\textbf{Methods} & \textbf{Identified Genes}\\
\hline
\hline
\begin{tabular}[c]{@{}l@{}}11 "real" hub genes identified at  intersection  of\\ hub genes produced by three ranking algorithms.\end{tabular} & \begin{tabular}[c]{@{}l@{}}NTRK2, COL14A1,  SCN4B,  ATP1A2,  SLC17A7, \\ SLIT3, COL7A1, CLDN4, CLEC3B, ADCYAP1R1, \\ MFAP4\end{tabular}                                                                                                                      \\ \hline \hline
{\color[HTML]{333333} \begin{tabular}[c]{@{}l@{}}Significant  genes identified through  the Pearson \\ Correlation FSM.\end{tabular}}         & \cellcolor[HTML]{FFFFFF}\begin{tabular}[c]{@{}l@{}}ABCA8,  ADAMTS3,  AFAP1-AS1,  AFF3,  ASTN1, \\ CCAT1,CLDN4,COL7A1,CYP24A1, DDR1, ESRP1, \\ EVPL,  FAM111B,  GALNT16,   KCNAB1,   PTGIS, \\ TMEM59L,  RAMP2 - AS1,  SCARNA5,  SLC22A4,\\ PTX3\end{tabular} \\ \hline
\hline
\begin{tabular}[c]{@{}l@{}}Significant  genes  identified  through  Recursive \\ Feature Elimination FSM\end{tabular}                         & \begin{tabular}[c]{@{}l@{}}ATP1B2, CLDN4, COLEC12,CRLF1,DDR1, EFHD1, \\ FAM111B, FHL1, GALNT16, JAM3, LSR,   NCEH1, \\ NRN1,PTGIS,SFRP1,SPINT2,SPON1,TSPAN7,TUB, \\ ST6GALNAC6\end{tabular}                                                                  \\ \hline
\end{tabular}
}

\caption{Identification of real hub genes by PPI network and significant genes by feature selection methods.}
\label{tab:hub}
\end{table}

\subsection{Identification of significant genes using feature selection methods}

By employing feature selection techniques to identify the gene sets that most effectively contribute to the prediction job, we also ascertained the features required to construct the ML models. The 147 DEGs were used in the application of the recursive feature elimination and pearson correlation feature selection procedures. To choose the most suitable group of features for both approaches, we experimented with a number of threshold values. Through defining the correlation scores above 0.9, we succeeded to get gene groups of 20 using Pearson correlation method. Furthermore, by setting the $n\_features\_to\_select$ to 20 and utilising the SVM approach as an $estimator$, we succeeded to get gene groups of 20 through the use of the recursive feature elimination method.
Table \ref{tab:hub} lists the significant genes that have been identified using both feature selection methods.

\subsection{Evaluating the prediction performance of ML models on independent dataset}

\begin{figure}[h]
   
    \centering
    \includegraphics[scale=0.3]{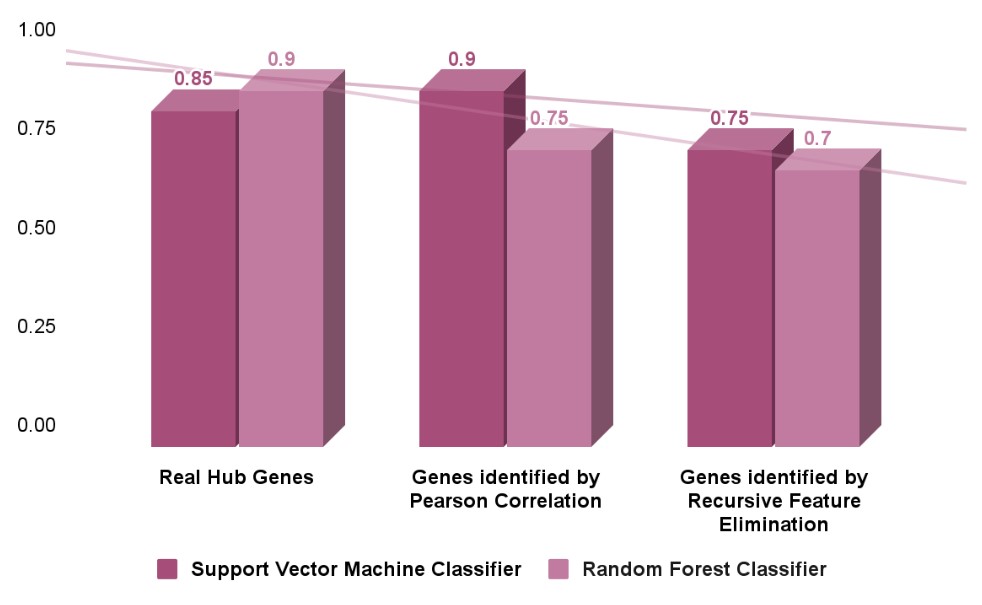}
    \caption{The graphical representation of accuracy scores of Support vector machine and Random Forest classification model based on real hub genes identified at the intersection of three hub gene ranking algorithm and genes identified by Pearson Correlation Method and Recursive Feature Elimination Method}
    \label{fig:acc}
\end{figure}

\begin{figure}[]
    \centering
    \subfigure[Real Hub gene subsets]{\includegraphics[scale=.4]{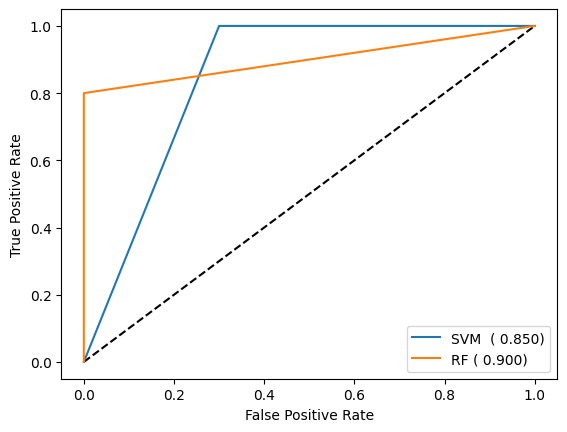}}
    \subfigure[Gene subsets identified by Pearson Correlation method]{\includegraphics[scale=.4]{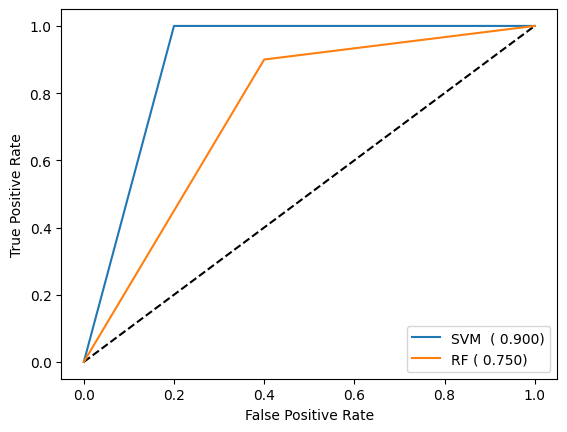}}
    \subfigure[Gene subsets identified by Recursive Feature Elimination method]{\includegraphics[scale=.4]{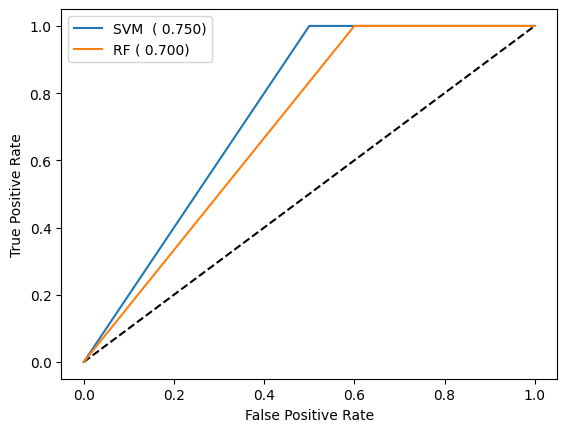}}
    \caption{The ROC curves of Support vector machine and Random Forest classification model based on real hub genes identified at the intersection of three hub gene ranking algorithm and genes identified by Pearson Correlation Method and Recursive Feature Elimination Method}
    \label{fig:roc}
\end{figure}

We assessed the variations in the prediction abilities of the two machine learning models, RF and SVM, on independent dataset(GSE139682) upon feeding them distinct feature sets of 11 'real' hub genes and of genes determined by the two feature selection techniques independently. Hence, we created three separate datasets for each biomarker identification methods. First dataset contains the 11 real hub genes as features; second and third dataset contains genes identified by pearson correlation and recursive feature elimination method respectively as features. Following that, these datasets were classified, and the classification abilities of individual feature selection techniques and classification algorithms were examined. The accuracy score of these investigations during the evaluation procedure are shown in table \ref{tab:accuracy} and in figure \ref{fig:acc}. Figure \ref{fig:roc}  shows the AUROC curve of the independent testing dataset for SVM, RF
classifiers with the highest performance for the identified different gene subsets. In all cases, the result shows that the subset of real hub genes outperformed than other genes subsets. On the other hand, the gene subset identified by Recursive Feature Elimination method provided worst performance. These findings imply that classifiers built using hub genes were able to attain good prediction performances.
\begin{table}[h]
\resizebox{\textwidth}{!}{
\begin{tabular}{|l|l|ll|}
\hline
   \multirow{2}{*} {\textbf{Models}   }                         & \multirow{2}{*}{\textbf{'Real' hub genes}} & \multicolumn{2}{|c|}{\textbf{\underline{Significant Genes Identified by feature selection methods}}}                  \\  
                                   &                                            & \multicolumn{1}{l|}{\textbf{Pearson Correlation Method}} & \textbf{Recursive Feature Elimination Method} \\ \hline
\multicolumn{1}{|l|}{\textbf{SVM}} & 0..85                                      & \multicolumn{1}{l|}{0.9}                                 & 0.75                                          \\ \hline
\multicolumn{1}{|l|}{\textbf{RF}}  & 0.9                                        & \multicolumn{1}{l|}{0.75}                                & 0.7                                           \\ \hline
\end{tabular}
}
\caption{The accuracy scores of Support vector machine and Random Forest classification model based on real hub genes identified at the intersection of three hub gene ranking algorithm and genes identified by Pearson Correlation Method and Recursive Feature Elimination Method}
\label{tab:accuracy}
\end{table}
\subsection{Validation of hub gene expression}
\begin{figure}[h]
    \centering
    \subfigure[NTRK2]{\includegraphics[scale=.35]{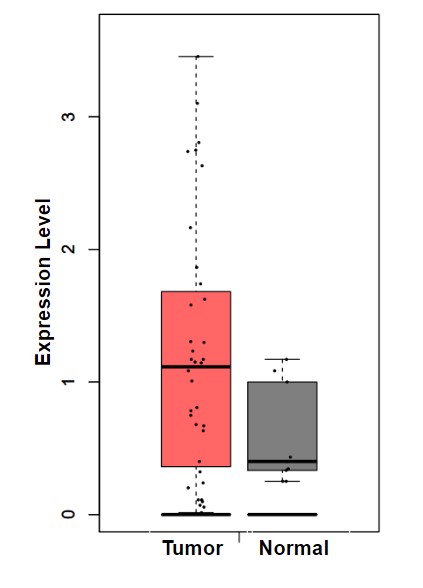}}
    \subfigure[COL14A1]{\includegraphics[scale=.35]{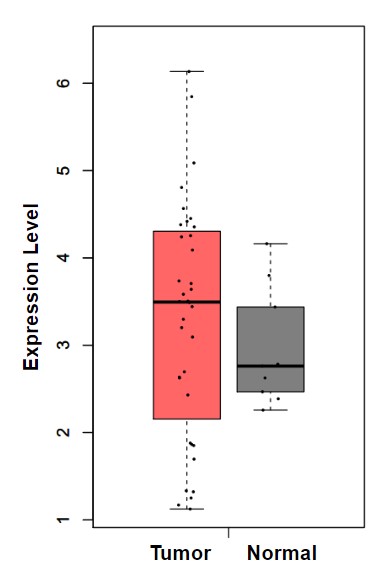}}
    \subfigure[SCN4B]{\includegraphics[scale=.35]{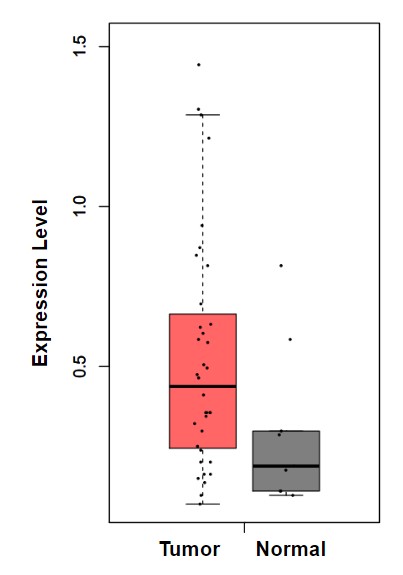}}
    \subfigure[ATP1A2]{\includegraphics[scale=.35]{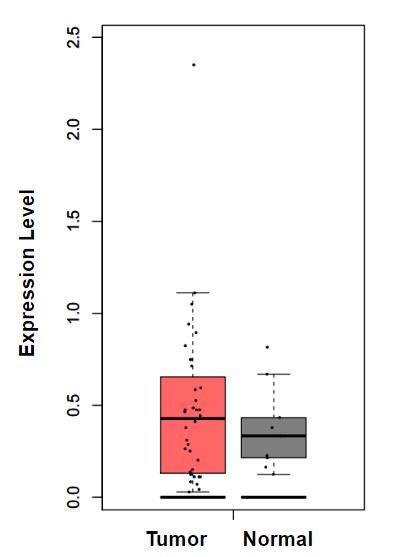}}
    \subfigure[SLC17A7]{\includegraphics[scale=.35]{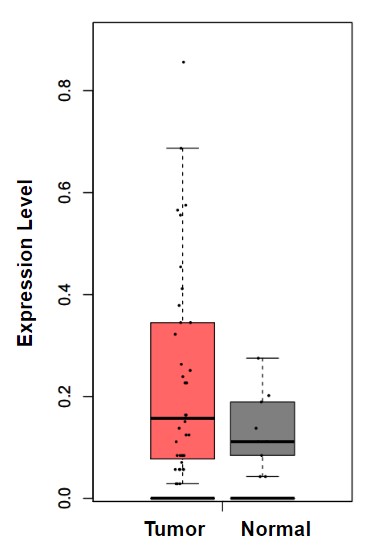}}
    \subfigure[SLIT3]{\includegraphics[scale=.35]{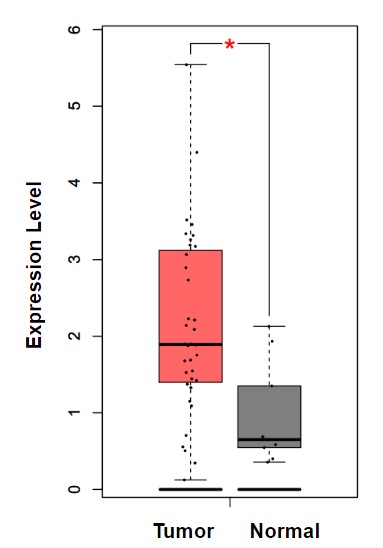}}
    \subfigure[COL7A1]{\includegraphics[scale=.35]{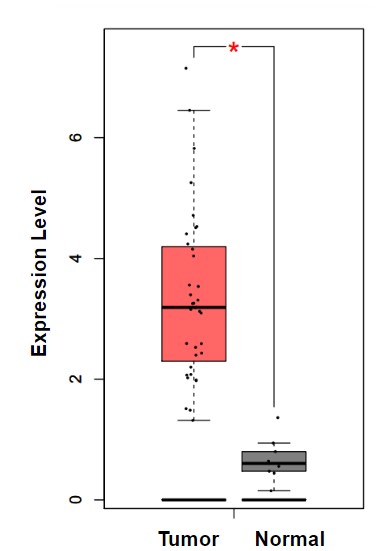}}
    \subfigure[CLDN4]{\includegraphics[scale=.35]{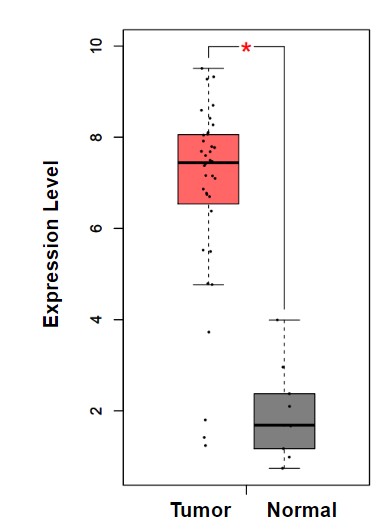}}
    \subfigure[CLEC3B]{\includegraphics[scale=.35]{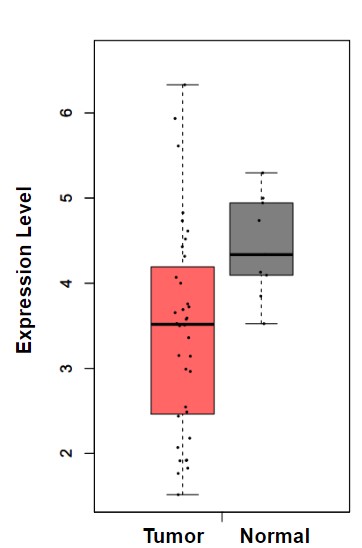}}
    \subfigure[ADCYAP1R1]{\includegraphics[scale=.35]{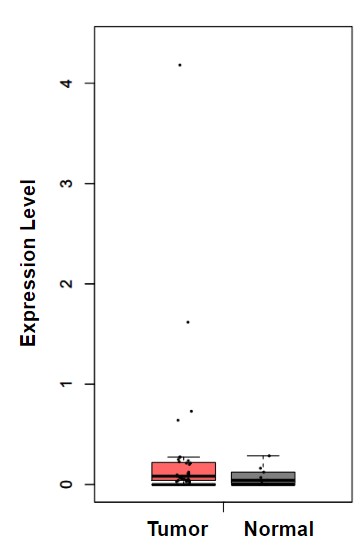}}
    \subfigure[MFAP4]{\includegraphics[scale=.35]{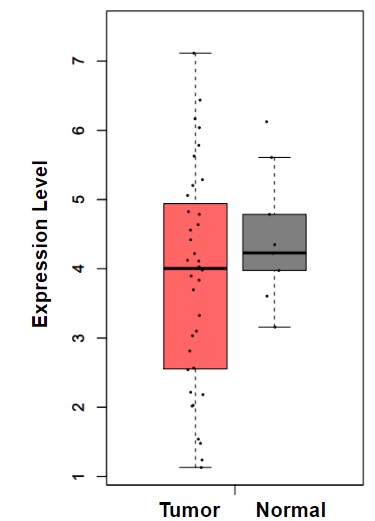}}
    \caption{Validation of hub gene expression: GEPIA files are used to create boxplots that display the hub gene expression in GBC patients and healthy controls.}
    \label{fig:gepia}
\end{figure}

Hub gene expression was confirmed using the GEPIA database, with a $P-value < 0.05$ and a $Log2FC > 1$ criterion. GEPIA box plots showed that all hub gene expressions in GBC patients were considerably upregulated in the SLIT3, COL7A1, CLDN4 (Figure: \ref{fig:gepia}). There are little statistical diference in other genes.

\section{Discussion}\label{sec4}

Gallbladder cancer (GBC) is the most common cause of disease among biliary tract neoplasms, comprising 80\% - 95\% \citep{huang2021worldwide}.  Based to GLOBOCAN's ( Global Cancer Observatory) 2020 cancer data \citep{khandelwal2017emerging}, GBC was the 24th most common cause of cancer worldwide in 2020, with over 115,949 newly reported cases. The amount of patients with a GBC diagnosis reached at nearly 84,695 that year because of the severe form of the cancer. The overall global incidence has been rising over recent years, despite regional variations in incidence rates. This trend is expected to continue as risk factors grow more common among populations \citep{liu2021long}. The goal and challenge for medical and scientific research has always been to identify the molecular mechanism and biomarkers associated with the onset and progression of gallbladder cancer. This research has significant research value in enhancing the diagnosis, efficacy of treatment, and prognosis survival of lung cancer patients. 

The current work aims to comprehensively identify potential genes and pathways associated with gallbladder cancer by bioinformatics analysis and machine learning methods. From the Gene Expression Omnibus database, gene expression data (expression profles GSE100363 and GSE139682) were acquired. Afterward, a total of 432 and 1800 potential DEGs were obtained from two different datasets. Of these, 146 were identified as common potential DEGs, with 39 of them being up-regulated genes and 107 being down-regulated genes related to gallbladder cancer. Next, in order to explore up-regulated and down-regulated genes, we carried out enrichment studies of GO analysis (three methods: CC, MF, and BP), pathway analysis using REACTOM database.

According to the GO 
 and REACTOME study, the overexpressed DEGs are mainly associated with biological processes like "cell adhesion," "epidermis development," and "keratinization," cellular components like "integral membrane" and "plasma membrane,", molecular functions like "structural molecule activity" and pathways like ”Signalling by Receptor Tyrosine Kinase.  Likewise, the investigation reveals that the downexpressed DEGs are primarily linked to ”cell differentiation”,
”nervous system development”, ”cell adhesion” for biological process, ”the plasma membrane”, ”integral component
of membrane” for cellular component, ”calcium ion binding”, and ”heparin binding” for molecular function and "Signal Transduction” for pathways.

In order to evaluate the interactional links, we also built a PPI network. After that, we obtained the top-15 hub genes for each of the three ranking algorithms: Closeness Centrality, MNC, and Degree. Surprisingly, we found that just 11 of the hub genes—which were regarded as "real" hub genes—were recognised by all three ranking techniques. Additionally, SLIT3, COL7A1, and CLDN4 hub gene expressions in GBC patients were significantly elevated, according to GEPIA box plots. Other genes show minimal statistical differences. 

Additionally, in this investigation, we employed feature selection methods including Recursive Feature Elimination (RFE) and Pearson Correlation to identify the significant DEGs that most effectively separate unhealthy samples from the healthy controls. Then, using the SVM and RF algorithms, the genuine hub genes and significant genes that were found using the feature selection method were trained on the GSE 100363 dataset to create a machine learning model. In order to validate the biomarkers, the model was lastly validated using the independent GSE 139682 dataset. Nonetheless, the outcomes showed that the subset of real hub genes outperformed the others, suggesting that these candidate genes could be used as
 potential biomarkers for GBC diagnosis.
\section{Conclusions}\label{sec5}

In this study, we analyzed gene expression data using bioinformatics and machine learning approach, which may supplement traditional
clinical prognostic factors, enabling clinicians to provide
more effective therapeutic intervention and personalized
treatment for gall bladder cancer. Here, We designed a computational method 
that exploits multiple hub gene ranking methods and feature selection methods with machine 
learning and bioinformatics to identify biomarkers. In this study, We have identified 11 key genes with diagnostic and prognostic values in GBC. We
identified these genes by using comprehensive bioinformatics and machine learning technology.

\backmatter




\noindent


\begin{flushleft}%




\end{flushleft}

\begin{appendices}






\end{appendices}


\bibliography{sn-bibliography}


\begin{thebibliography}{35}
\ifx \bisbn   \undefined \def \bisbn  #1{ISBN #1}\fi
\ifx \binits  \undefined \def \binits#1{#1}\fi
\ifx \bauthor  \undefined \def \bauthor#1{#1}\fi
\ifx \batitle  \undefined \def \batitle#1{#1}\fi
\ifx \bjtitle  \undefined \def \bjtitle#1{#1}\fi
\ifx \bvolume  \undefined \def \bvolume#1{\textbf{#1}}\fi
\ifx \byear  \undefined \def \byear#1{#1}\fi
\ifx \bissue  \undefined \def \bissue#1{#1}\fi
\ifx \bfpage  \undefined \def \bfpage#1{#1}\fi
\ifx \blpage  \undefined \def \blpage #1{#1}\fi
\ifx \burl  \undefined \def \burl#1{\textsf{#1}}\fi
\ifx \doiurl  \undefined \def \doiurl#1{\url{https://doi.org/#1}}\fi
\ifx \betal  \undefined \def \betal{\textit{et al.}}\fi
\ifx \binstitute  \undefined \def \binstitute#1{#1}\fi
\ifx \binstitutionaled  \undefined \def \binstitutionaled#1{#1}\fi
\ifx \bctitle  \undefined \def \bctitle#1{#1}\fi
\ifx \beditor  \undefined \def \beditor#1{#1}\fi
\ifx \bpublisher  \undefined \def \bpublisher#1{#1}\fi
\ifx \bbtitle  \undefined \def \bbtitle#1{#1}\fi
\ifx \bedition  \undefined \def \bedition#1{#1}\fi
\ifx \bseriesno  \undefined \def \bseriesno#1{#1}\fi
\ifx \blocation  \undefined \def \blocation#1{#1}\fi
\ifx \bsertitle  \undefined \def \bsertitle#1{#1}\fi
\ifx \bsnm \undefined \def \bsnm#1{#1}\fi
\ifx \bsuffix \undefined \def \bsuffix#1{#1}\fi
\ifx \bparticle \undefined \def \bparticle#1{#1}\fi
\ifx \barticle \undefined \def \barticle#1{#1}\fi
\bibcommenthead
\ifx \bconfdate \undefined \def \bconfdate #1{#1}\fi
\ifx \botherref \undefined \def \botherref #1{#1}\fi
\ifx \url \undefined \def \url#1{\textsf{#1}}\fi
\ifx \bchapter \undefined \def \bchapter#1{#1}\fi
\ifx \bbook \undefined \def \bbook#1{#1}\fi
\ifx \bcomment \undefined \def \bcomment#1{#1}\fi
\ifx \oauthor \undefined \def \oauthor#1{#1}\fi
\ifx \citeauthoryear \undefined \def \citeauthoryear#1{#1}\fi
\ifx \endbibitem  \undefined \def \endbibitem {}\fi
\ifx \bconflocation  \undefined \def \bconflocation#1{#1}\fi
\ifx \arxivurl  \undefined \def \arxivurl#1{\textsf{#1}}\fi
\csname PreBibitemsHook\endcsname

\bibitem[\protect\citeauthoryear{Hundal and Shaffer}{2014}]{hundal2014gallbladder}
\begin{botherref}
\oauthor{\bsnm{Hundal}, \binits{R.}},
\oauthor{\bsnm{Shaffer}, \binits{E.A.}}:
Gallbladder cancer: epidemiology and outcome.
Clinical epidemiology,
99--109
(2014)
\end{botherref}
\endbibitem

\bibitem[\protect\citeauthoryear{Raki{\'c} et~al.}{2014}]{rakic2014gallbladder}
\begin{barticle}
\bauthor{\bsnm{Raki{\'c}}, \binits{M.}},
\bauthor{\bsnm{Patrlj}, \binits{L.}},
\bauthor{\bsnm{Kopljar}, \binits{M.}},
\bauthor{\bsnm{Kli{\v{c}}ek}, \binits{R.}},
\bauthor{\bsnm{Kolovrat}, \binits{M.}},
\bauthor{\bsnm{Loncar}, \binits{B.}},
\bauthor{\bsnm{Busic}, \binits{Z.}}:
\batitle{Gallbladder cancer}.
\bjtitle{Hepatobiliary surgery and nutrition}
\bvolume{3}(\bissue{5}),
\bfpage{221}
(\byear{2014})
\end{barticle}
\endbibitem

\bibitem[\protect\citeauthoryear{Siegel et~al.}{2021}]{siegel2021cancer}
\begin{barticle}
\bauthor{\bsnm{Siegel}, \binits{R.L.}},
\bauthor{\bsnm{Miller}, \binits{K.D.}},
\bauthor{\bsnm{Fuchs}, \binits{H.E.}},
\bauthor{\bsnm{Jemal}, \binits{A.}}, \betal:
\batitle{Cancer statistics, 2021}.
\bjtitle{Ca Cancer J Clin}
\bvolume{71}(\bissue{1}),
\bfpage{7}--\blpage{33}
(\byear{2021})
\end{barticle}
\endbibitem

\bibitem[\protect\citeauthoryear{Valle et~al.}{2021}]{valle2021biliary}
\begin{barticle}
\bauthor{\bsnm{Valle}, \binits{J.W.}},
\bauthor{\bsnm{Kelley}, \binits{R.K.}},
\bauthor{\bsnm{Nervi}, \binits{B.}},
\bauthor{\bsnm{Oh}, \binits{D.-Y.}},
\bauthor{\bsnm{Zhu}, \binits{A.X.}}:
\batitle{Biliary tract cancer}.
\bjtitle{The Lancet}
\bvolume{397}(\bissue{10272}),
\bfpage{428}--\blpage{444}
(\byear{2021})
\end{barticle}
\endbibitem

\bibitem[\protect\citeauthoryear{Gourgiotis et~al.}{2008}]{gourgiotis2008gallbladder}
\begin{barticle}
\bauthor{\bsnm{Gourgiotis}, \binits{S.}},
\bauthor{\bsnm{Kocher}, \binits{H.M.}},
\bauthor{\bsnm{Solaini}, \binits{L.}},
\bauthor{\bsnm{Yarollahi}, \binits{A.}},
\bauthor{\bsnm{Tsiambas}, \binits{E.}},
\bauthor{\bsnm{Salemis}, \binits{N.S.}}:
\batitle{Gallbladder cancer}.
\bjtitle{The American Journal of Surgery}
\bvolume{196}(\bissue{2}),
\bfpage{252}--\blpage{264}
(\byear{2008})
\end{barticle}
\endbibitem

\bibitem[\protect\citeauthoryear{Chun et~al.}{2018}]{chun2018ajcc}
\begin{barticle}
\bauthor{\bsnm{Chun}, \binits{Y.S.}},
\bauthor{\bsnm{Pawlik}, \binits{T.M.}},
\bauthor{\bsnm{Vauthey}, \binits{J.-N.}}:
\batitle{of the ajcc cancer staging manual: pancreas and hepatobiliary cancers}.
\bjtitle{Annals of surgical oncology}
\bvolume{25},
\bfpage{845}--\blpage{847}
(\byear{2018})
\end{barticle}
\endbibitem

\bibitem[\protect\citeauthoryear{Mantripragada et~al.}{2017}]{mantripragada2017adjuvant}
\begin{barticle}
\bauthor{\bsnm{Mantripragada}, \binits{K.C.}},
\bauthor{\bsnm{Hamid}, \binits{F.}},
\bauthor{\bsnm{Shafqat}, \binits{H.}},
\bauthor{\bsnm{Olszewski}, \binits{A.J.}}:
\batitle{Adjuvant therapy for resected gallbladder cancer: analysis of the national cancer data base}.
\bjtitle{Journal of the National Cancer Institute}
\bvolume{109}(\bissue{2}),
\bfpage{202}
(\byear{2017})
\end{barticle}
\endbibitem

\bibitem[\protect\citeauthoryear{Chen et~al.}{2019}]{chen2019development}
\begin{botherref}
\oauthor{\bsnm{Chen}, \binits{M.}},
\oauthor{\bsnm{Cao}, \binits{J.}},
\oauthor{\bsnm{Bai}, \binits{Y.}},
\oauthor{\bsnm{Tong}, \binits{C.}},
\oauthor{\bsnm{Lin}, \binits{J.}},
\oauthor{\bsnm{Jindal}, \binits{V.}},
\oauthor{\bsnm{Barchi}, \binits{L.C.}},
\oauthor{\bsnm{Nadalin}, \binits{S.}},
\oauthor{\bsnm{Yang}, \binits{S.X.}},
\oauthor{\bsnm{Pesce}, \binits{A.}}, et al.:
Development and validation of a nomogram for early detection of malignant gallbladder lesions.
Clinical and translational gastroenterology
\textbf{10}(10)
(2019)
\end{botherref}
\endbibitem

\bibitem[\protect\citeauthoryear{Tauriello et~al.}{2018}]{tauriello2018tgfbeta}
\begin{barticle}
\bauthor{\bsnm{Tauriello}, \binits{D.V.}},
\bauthor{\bsnm{Palomo-Ponce}, \binits{S.}},
\bauthor{\bsnm{Stork}, \binits{D.}},
\bauthor{\bsnm{Berenguer-Llergo}, \binits{A.}},
\bauthor{\bsnm{Badia-Ramentol}, \binits{J.}},
\bauthor{\bsnm{Iglesias}, \binits{M.}},
\bauthor{\bsnm{Sevillano}, \binits{M.}},
\bauthor{\bsnm{Ibiza}, \binits{S.}},
\bauthor{\bsnm{Ca{\~n}ellas}, \binits{A.}},
\bauthor{\bsnm{Hernando-Momblona}, \binits{X.}}, \betal:
\batitle{Tgf$\beta$ drives immune evasion in genetically reconstituted colon cancer metastasis}.
\bjtitle{Nature}
\bvolume{554}(\bissue{7693}),
\bfpage{538}--\blpage{543}
(\byear{2018})
\end{barticle}
\endbibitem

\bibitem[\protect\citeauthoryear{Wu et~al.}{2019}]{wu2019epithelial}
\begin{barticle}
\bauthor{\bsnm{Wu}, \binits{M.-J.}},
\bauthor{\bsnm{Chen}, \binits{Y.-S.}},
\bauthor{\bsnm{Kim}, \binits{M.R.}},
\bauthor{\bsnm{Chang}, \binits{C.-C.}},
\bauthor{\bsnm{Gampala}, \binits{S.}},
\bauthor{\bsnm{Zhang}, \binits{Y.}},
\bauthor{\bsnm{Wang}, \binits{Y.}},
\bauthor{\bsnm{Chang}, \binits{C.-Y.}},
\bauthor{\bsnm{Yang}, \binits{J.-Y.}},
\bauthor{\bsnm{Chang}, \binits{C.-J.}}:
\batitle{Epithelial-mesenchymal transition directs stem cell polarity via regulation of mitofusin}.
\bjtitle{Cell metabolism}
\bvolume{29}(\bissue{4}),
\bfpage{993}--\blpage{1002}
(\byear{2019})
\end{barticle}
\endbibitem

\bibitem[\protect\citeauthoryear{Zheng et~al.}{2018}]{zheng2018elf3}
\begin{barticle}
\bauthor{\bsnm{Zheng}, \binits{L.}},
\bauthor{\bsnm{Xu}, \binits{M.}},
\bauthor{\bsnm{Xu}, \binits{J.}},
\bauthor{\bsnm{Wu}, \binits{K.}},
\bauthor{\bsnm{Fang}, \binits{Q.}},
\bauthor{\bsnm{Liang}, \binits{Y.}},
\bauthor{\bsnm{Zhou}, \binits{S.}},
\bauthor{\bsnm{Cen}, \binits{D.}},
\bauthor{\bsnm{Ji}, \binits{L.}},
\bauthor{\bsnm{Han}, \binits{W.}}, \betal:
\batitle{Elf3 promotes epithelial--mesenchymal transition by protecting zeb1 from mir-141-3p-mediated silencing in hepatocellular carcinoma}.
\bjtitle{Cell death \& disease}
\bvolume{9}(\bissue{3}),
\bfpage{387}
(\byear{2018})
\end{barticle}
\endbibitem

\bibitem[\protect\citeauthoryear{Civenni et~al.}{2019}]{civenni2019epigenetic}
\begin{barticle}
\bauthor{\bsnm{Civenni}, \binits{G.}},
\bauthor{\bsnm{Bosotti}, \binits{R.}},
\bauthor{\bsnm{Timpanaro}, \binits{A.}},
\bauthor{\bsnm{Vazquez}, \binits{R.}},
\bauthor{\bsnm{Merulla}, \binits{J.}},
\bauthor{\bsnm{Pandit}, \binits{S.}},
\bauthor{\bsnm{Rossi}, \binits{S.}},
\bauthor{\bsnm{Albino}, \binits{D.}},
\bauthor{\bsnm{Allegrini}, \binits{S.}},
\bauthor{\bsnm{Mitra}, \binits{A.}}, \betal:
\batitle{Epigenetic control of mitochondrial fission enables self-renewal of stem-like tumor cells in human prostate cancer}.
\bjtitle{Cell metabolism}
\bvolume{30}(\bissue{2}),
\bfpage{303}--\blpage{318}
(\byear{2019})
\end{barticle}
\endbibitem

\bibitem[\protect\citeauthoryear{Raphael et~al.}{2017}]{raphael2017integrated}
\begin{barticle}
\bauthor{\bsnm{Raphael}, \binits{B.J.}},
\bauthor{\bsnm{Hruban}, \binits{R.H.}},
\bauthor{\bsnm{Aguirre}, \binits{A.J.}},
\bauthor{\bsnm{Moffitt}, \binits{R.A.}},
\bauthor{\bsnm{Yeh}, \binits{J.J.}},
\bauthor{\bsnm{Stewart}, \binits{C.}},
\bauthor{\bsnm{Robertson}, \binits{A.G.}},
\bauthor{\bsnm{Cherniack}, \binits{A.D.}},
\bauthor{\bsnm{Gupta}, \binits{M.}},
\bauthor{\bsnm{Getz}, \binits{G.}}, \betal:
\batitle{Integrated genomic characterization of pancreatic ductal adenocarcinoma}.
\bjtitle{Cancer cell}
\bvolume{32}(\bissue{2}),
\bfpage{185}--\blpage{203}
(\byear{2017})
\end{barticle}
\endbibitem

\bibitem[\protect\citeauthoryear{Kinde et~al.}{2013}]{kinde2013evaluation}
\begin{barticle}
\bauthor{\bsnm{Kinde}, \binits{I.}},
\bauthor{\bsnm{Bettegowda}, \binits{C.}},
\bauthor{\bsnm{Wang}, \binits{Y.}},
\bauthor{\bsnm{Wu}, \binits{J.}},
\bauthor{\bsnm{Agrawal}, \binits{N.}},
\bauthor{\bsnm{Shih}, \binits{I.-M.}},
\bauthor{\bsnm{Kurman}, \binits{R.}},
\bauthor{\bsnm{Dao}, \binits{F.}},
\bauthor{\bsnm{Levine}, \binits{D.A.}},
\bauthor{\bsnm{Giuntoli}, \binits{R.}}, \betal:
\batitle{Evaluation of dna from the papanicolaou test to detect ovarian and endometrial cancers}.
\bjtitle{Science translational medicine}
\bvolume{5}(\bissue{167}),
\bfpage{167}--\blpage{41674}
(\byear{2013})
\end{barticle}
\endbibitem

\bibitem[\protect\citeauthoryear{Kulasingam and Diamandis}{2008}]{kulasingam2008strategies}
\begin{barticle}
\bauthor{\bsnm{Kulasingam}, \binits{V.}},
\bauthor{\bsnm{Diamandis}, \binits{E.P.}}:
\batitle{Strategies for discovering novel cancer biomarkers through utilization of emerging technologies}.
\bjtitle{Nature clinical practice Oncology}
\bvolume{5}(\bissue{10}),
\bfpage{588}--\blpage{599}
(\byear{2008})
\end{barticle}
\endbibitem

\bibitem[\protect\citeauthoryear{Auslander et~al.}{2021}]{auslander2021incorporating}
\begin{barticle}
\bauthor{\bsnm{Auslander}, \binits{N.}},
\bauthor{\bsnm{Gussow}, \binits{A.B.}},
\bauthor{\bsnm{Koonin}, \binits{E.V.}}:
\batitle{Incorporating machine learning into established bioinformatics frameworks}.
\bjtitle{International journal of molecular sciences}
\bvolume{22}(\bissue{6}),
\bfpage{2903}
(\byear{2021})
\end{barticle}
\endbibitem

\bibitem[\protect\citeauthoryear{Barrett et~al.}{2012}]{barrett2012ncbi}
\begin{barticle}
\bauthor{\bsnm{Barrett}, \binits{T.}},
\bauthor{\bsnm{Wilhite}, \binits{S.E.}},
\bauthor{\bsnm{Ledoux}, \binits{P.}},
\bauthor{\bsnm{Evangelista}, \binits{C.}},
\bauthor{\bsnm{Kim}, \binits{I.F.}},
\bauthor{\bsnm{Tomashevsky}, \binits{M.}},
\bauthor{\bsnm{Marshall}, \binits{K.A.}},
\bauthor{\bsnm{Phillippy}, \binits{K.H.}},
\bauthor{\bsnm{Sherman}, \binits{P.M.}},
\bauthor{\bsnm{Holko}, \binits{M.}}, \betal:
\batitle{Ncbi geo: archive for functional genomics data sets—update}.
\bjtitle{Nucleic acids research}
\bvolume{41}(\bissue{D1}),
\bfpage{991}--\blpage{995}
(\byear{2012})
\end{barticle}
\endbibitem

\bibitem[\protect\citeauthoryear{Kong et~al.}{2023}]{kong2023screening}
\begin{barticle}
\bauthor{\bsnm{Kong}, \binits{X.}},
\bauthor{\bsnm{Wang}, \binits{C.}},
\bauthor{\bsnm{Wu}, \binits{Q.}},
\bauthor{\bsnm{Wang}, \binits{Z.}},
\bauthor{\bsnm{Han}, \binits{Y.}},
\bauthor{\bsnm{Teng}, \binits{J.}},
\bauthor{\bsnm{Qi}, \binits{X.}}:
\batitle{Screening and identification of key biomarkers of depression using bioinformatics}.
\bjtitle{Scientific Reports}
\bvolume{13}(\bissue{1}),
\bfpage{4180}
(\byear{2023})
\end{barticle}
\endbibitem

\bibitem[\protect\citeauthoryear{Xu et~al.}{2016}]{xu2016identification}
\begin{barticle}
\bauthor{\bsnm{Xu}, \binits{Z.}},
\bauthor{\bsnm{Zhou}, \binits{Y.}},
\bauthor{\bsnm{Cao}, \binits{Y.}},
\bauthor{\bsnm{Dinh}, \binits{T.L.A.}},
\bauthor{\bsnm{Wan}, \binits{J.}},
\bauthor{\bsnm{Zhao}, \binits{M.}}:
\batitle{Identification of candidate biomarkers and analysis of prognostic values in ovarian cancer by integrated bioinformatics analysis}.
\bjtitle{Medical oncology}
\bvolume{33},
\bfpage{1}--\blpage{8}
(\byear{2016})
\end{barticle}
\endbibitem

\bibitem[\protect\citeauthoryear{Oliveros}{2007}]{oliveros2007venny}
\begin{botherref}
\oauthor{\bsnm{Oliveros}, \binits{J.C.}}:
Venny. an interactive tool for comparing lists with venn diagrams.
http://bioinfogp. cnb. csic. es/tools/venny/index. html
(2007)
\end{botherref}
\endbibitem

\bibitem[\protect\citeauthoryear{Reimand et~al.}{2019}]{reimand2019pathway}
\begin{barticle}
\bauthor{\bsnm{Reimand}, \binits{J.}},
\bauthor{\bsnm{Isserlin}, \binits{R.}},
\bauthor{\bsnm{Voisin}, \binits{V.}},
\bauthor{\bsnm{Kucera}, \binits{M.}},
\bauthor{\bsnm{Tannus-Lopes}, \binits{C.}},
\bauthor{\bsnm{Rostamianfar}, \binits{A.}},
\bauthor{\bsnm{Wadi}, \binits{L.}},
\bauthor{\bsnm{Meyer}, \binits{M.}},
\bauthor{\bsnm{Wong}, \binits{J.}},
\bauthor{\bsnm{Xu}, \binits{C.}}, \betal:
\batitle{Pathway enrichment analysis and visualization of omics data using g: Profiler, gsea, cytoscape and enrichmentmap}.
\bjtitle{Nature protocols}
\bvolume{14}(\bissue{2}),
\bfpage{482}--\blpage{517}
(\byear{2019})
\end{barticle}
\endbibitem

\bibitem[\protect\citeauthoryear{Ashburner et~al.}{2000}]{ashburner2000gene}
\begin{barticle}
\bauthor{\bsnm{Ashburner}, \binits{M.}},
\bauthor{\bsnm{Ball}, \binits{C.A.}},
\bauthor{\bsnm{Blake}, \binits{J.A.}},
\bauthor{\bsnm{Botstein}, \binits{D.}},
\bauthor{\bsnm{Butler}, \binits{H.}},
\bauthor{\bsnm{Cherry}, \binits{J.M.}},
\bauthor{\bsnm{Davis}, \binits{A.P.}},
\bauthor{\bsnm{Dolinski}, \binits{K.}},
\bauthor{\bsnm{Dwight}, \binits{S.S.}},
\bauthor{\bsnm{Eppig}, \binits{J.T.}}, \betal:
\batitle{Gene ontology: tool for the unification of biology}.
\bjtitle{Nature genetics}
\bvolume{25}(\bissue{1}),
\bfpage{25}--\blpage{29}
(\byear{2000})
\end{barticle}
\endbibitem

\bibitem[\protect\citeauthoryear{Croft et~al.}{2010}]{croft2010reactome}
\begin{barticle}
\bauthor{\bsnm{Croft}, \binits{D.}},
\bauthor{\bsnm{O’kelly}, \binits{G.}},
\bauthor{\bsnm{Wu}, \binits{G.}},
\bauthor{\bsnm{Haw}, \binits{R.}},
\bauthor{\bsnm{Gillespie}, \binits{M.}},
\bauthor{\bsnm{Matthews}, \binits{L.}},
\bauthor{\bsnm{Caudy}, \binits{M.}},
\bauthor{\bsnm{Garapati}, \binits{P.}},
\bauthor{\bsnm{Gopinath}, \binits{G.}},
\bauthor{\bsnm{Jassal}, \binits{B.}}, \betal:
\batitle{Reactome: a database of reactions, pathways and biological processes}.
\bjtitle{Nucleic acids research}
\bvolume{39}(\bissue{suppl\_1}),
\bfpage{691}--\blpage{697}
(\byear{2010})
\end{barticle}
\endbibitem

\bibitem[\protect\citeauthoryear{Huang et~al.}{2007}]{huang2007david}
\begin{barticle}
\bauthor{\bsnm{Huang}, \binits{D.W.}},
\bauthor{\bsnm{Sherman}, \binits{B.T.}},
\bauthor{\bsnm{Tan}, \binits{Q.}},
\bauthor{\bsnm{Collins}, \binits{J.R.}},
\bauthor{\bsnm{Alvord}, \binits{W.G.}},
\bauthor{\bsnm{Roayaei}, \binits{J.}},
\bauthor{\bsnm{Stephens}, \binits{R.}},
\bauthor{\bsnm{Baseler}, \binits{M.W.}},
\bauthor{\bsnm{Lane}, \binits{H.C.}},
\bauthor{\bsnm{Lempicki}, \binits{R.A.}}:
\batitle{The david gene functional classification tool: a novel biological module-centric algorithm to functionally analyze large gene lists}.
\bjtitle{Genome biology}
\bvolume{8}(\bissue{9}),
\bfpage{1}--\blpage{16}
(\byear{2007})
\end{barticle}
\endbibitem

\bibitem[\protect\citeauthoryear{Franceschini et~al.}{2012}]{franceschini2012string}
\begin{barticle}
\bauthor{\bsnm{Franceschini}, \binits{A.}},
\bauthor{\bsnm{Szklarczyk}, \binits{D.}},
\bauthor{\bsnm{Frankild}, \binits{S.}},
\bauthor{\bsnm{Kuhn}, \binits{M.}},
\bauthor{\bsnm{Simonovic}, \binits{M.}},
\bauthor{\bsnm{Roth}, \binits{A.}},
\bauthor{\bsnm{Lin}, \binits{J.}},
\bauthor{\bsnm{Minguez}, \binits{P.}},
\bauthor{\bsnm{Bork}, \binits{P.}},
\bauthor{\bsnm{Von~Mering}, \binits{C.}}, \betal:
\batitle{String v9. 1: protein-protein interaction networks, with increased coverage and integration}.
\bjtitle{Nucleic acids research}
\bvolume{41}(\bissue{D1}),
\bfpage{808}--\blpage{815}
(\byear{2012})
\end{barticle}
\endbibitem

\bibitem[\protect\citeauthoryear{Smoot et~al.}{2011}]{smoot2011cytoscape}
\begin{barticle}
\bauthor{\bsnm{Smoot}, \binits{M.E.}},
\bauthor{\bsnm{Ono}, \binits{K.}},
\bauthor{\bsnm{Ruscheinski}, \binits{J.}},
\bauthor{\bsnm{Wang}, \binits{P.-L.}},
\bauthor{\bsnm{Ideker}, \binits{T.}}:
\batitle{Cytoscape 2.8: new features for data integration and network visualization}.
\bjtitle{Bioinformatics}
\bvolume{27}(\bissue{3}),
\bfpage{431}--\blpage{432}
(\byear{2011})
\end{barticle}
\endbibitem

\bibitem[\protect\citeauthoryear{Kohl et~al.}{2011}]{kohl2011cytoscape}
\begin{botherref}
\oauthor{\bsnm{Kohl}, \binits{M.}},
\oauthor{\bsnm{Wiese}, \binits{S.}},
\oauthor{\bsnm{Warscheid}, \binits{B.}}:
Cytoscape: software for visualization and analysis of biological networks.
Data mining in proteomics: from standards to applications,
291--303
(2011)
\end{botherref}
\endbibitem

\bibitem[\protect\citeauthoryear{Chin et~al.}{2014}]{chin2014cytohubba}
\begin{barticle}
\bauthor{\bsnm{Chin}, \binits{C.-H.}},
\bauthor{\bsnm{Chen}, \binits{S.-H.}},
\bauthor{\bsnm{Wu}, \binits{H.-H.}},
\bauthor{\bsnm{Ho}, \binits{C.-W.}},
\bauthor{\bsnm{Ko}, \binits{M.-T.}},
\bauthor{\bsnm{Lin}, \binits{C.-Y.}}:
\batitle{cytohubba: identifying hub objects and sub-networks from complex interactome}.
\bjtitle{BMC systems biology}
\bvolume{8}(\bissue{4}),
\bfpage{1}--\blpage{7}
(\byear{2014})
\end{barticle}
\endbibitem

\bibitem[\protect\citeauthoryear{Breiman}{2001}]{breiman2001random}
\begin{barticle}
\bauthor{\bsnm{Breiman}, \binits{L.}}:
\batitle{Random forests}.
\bjtitle{Machine learning}
\bvolume{45},
\bfpage{5}--\blpage{32}
(\byear{2001})
\end{barticle}
\endbibitem

\bibitem[\protect\citeauthoryear{Khatun et~al.}{2023}]{khatun2023cancer}
\begin{barticle}
\bauthor{\bsnm{Khatun}, \binits{R.}},
\bauthor{\bsnm{Akter}, \binits{M.}},
\bauthor{\bsnm{Islam}, \binits{M.M.}},
\bauthor{\bsnm{Uddin}, \binits{M.A.}},
\bauthor{\bsnm{Talukder}, \binits{M.A.}},
\bauthor{\bsnm{Kamruzzaman}, \binits{J.}},
\bauthor{\bsnm{Azad}, \binits{A.}},
\bauthor{\bsnm{Paul}, \binits{B.K.}},
\bauthor{\bsnm{Almoyad}, \binits{M.A.A.}},
\bauthor{\bsnm{Aryal}, \binits{S.}}, \betal:
\batitle{Cancer classification utilizing voting classifier with ensemble feature selection method and transcriptomic data}.
\bjtitle{Genes}
\bvolume{14}(\bissue{9}),
\bfpage{1802}
(\byear{2023})
\end{barticle}
\endbibitem

\bibitem[\protect\citeauthoryear{Fakoor et~al.}{2013}]{fakoor2013using}
\begin{bchapter}
\bauthor{\bsnm{Fakoor}, \binits{R.}},
\bauthor{\bsnm{Ladhak}, \binits{F.}},
\bauthor{\bsnm{Nazi}, \binits{A.}},
\bauthor{\bsnm{Huber}, \binits{M.}}:
\bctitle{Using deep learning to enhance cancer diagnosis and classification}.
In: \bbtitle{Proceedings of the International Conference on Machine Learning},
vol. \bseriesno{28},
pp. \bfpage{3937}--\blpage{3949}
(\byear{2013}).
\bcomment{ACM New York, NY, USA}
\end{bchapter}
\endbibitem

\bibitem[\protect\citeauthoryear{Tang et~al.}{2017}]{tang2017gepia}
\begin{barticle}
\bauthor{\bsnm{Tang}, \binits{Z.}},
\bauthor{\bsnm{Li}, \binits{C.}},
\bauthor{\bsnm{Kang}, \binits{B.}},
\bauthor{\bsnm{Gao}, \binits{G.}},
\bauthor{\bsnm{Li}, \binits{C.}},
\bauthor{\bsnm{Zhang}, \binits{Z.}}:
\batitle{Gepia: a web server for cancer and normal gene expression profiling and interactive analyses}.
\bjtitle{Nucleic acids research}
\bvolume{45}(\bissue{W1}),
\bfpage{98}--\blpage{102}
(\byear{2017})
\end{barticle}
\endbibitem

\bibitem[\protect\citeauthoryear{Huang et~al.}{2021}]{huang2021worldwide}
\begin{barticle}
\bauthor{\bsnm{Huang}, \binits{J.}},
\bauthor{\bsnm{Patel}, \binits{H.K.}},
\bauthor{\bsnm{Boakye}, \binits{D.}},
\bauthor{\bsnm{Chandrasekar}, \binits{V.T.}},
\bauthor{\bsnm{Koulaouzidis}, \binits{A.}},
\bauthor{\bsnm{Lucero-Prisno~III}, \binits{D.E.}},
\bauthor{\bsnm{Ngai}, \binits{C.H.}},
\bauthor{\bsnm{Pun}, \binits{C.N.}},
\bauthor{\bsnm{Bai}, \binits{Y.}},
\bauthor{\bsnm{Lok}, \binits{V.}}, \betal:
\batitle{Worldwide distribution, associated factors, and trends of gallbladder cancer: a global country-level analysis}.
\bjtitle{Cancer letters}
\bvolume{521},
\bfpage{238}--\blpage{251}
(\byear{2021})
\end{barticle}
\endbibitem

\bibitem[\protect\citeauthoryear{Khandelwal et~al.}{2017}]{khandelwal2017emerging}
\begin{barticle}
\bauthor{\bsnm{Khandelwal}, \binits{A.}},
\bauthor{\bsnm{Malhotra}, \binits{A.}},
\bauthor{\bsnm{Jain}, \binits{M.}},
\bauthor{\bsnm{Vasquez}, \binits{K.M.}},
\bauthor{\bsnm{Jain}, \binits{A.}}:
\batitle{The emerging role of long non-coding rna in gallbladder cancer pathogenesis}.
\bjtitle{Biochimie}
\bvolume{132},
\bfpage{152}--\blpage{160}
(\byear{2017})
\end{barticle}
\endbibitem

\bibitem[\protect\citeauthoryear{Liu et~al.}{2021}]{liu2021long}
\begin{barticle}
\bauthor{\bsnm{Liu}, \binits{Y.}},
\bauthor{\bsnm{Ding}, \binits{W.}},
\bauthor{\bsnm{Yu}, \binits{W.}},
\bauthor{\bsnm{Zhang}, \binits{Y.}},
\bauthor{\bsnm{Ao}, \binits{X.}},
\bauthor{\bsnm{Wang}, \binits{J.}}:
\batitle{Long non-coding rnas: Biogenesis, functions, and clinical significance in gastric cancer}.
\bjtitle{Molecular Therapy-Oncolytics}
\bvolume{23},
\bfpage{458}--\blpage{476}
(\byear{2021})
\end{barticle}
\endbibitem

\end{thebibliography}

\end{document}